\documentclass[showpacs,twocolumn,floats]{revtex4}
\usepackage[latin1]{inputenc}
\usepackage[english]{babel}
\usepackage{graphicx}
\usepackage{psfrag}
\usepackage{amssymb}
\usepackage{amsmath}
\usepackage{amscd}
\usepackage{eucal}
\usepackage{color}
\usepackage{bm}

\newcommand{\rmd}{{\rm d}}
\newcommand{\rmi}{{\rm i}}

\newcommand{\de}{\delta}

\newcommand{\tE}{\tau_{\rm E}}

\newcommand{\tD}{\tau_{\rm D,S}}

\begin{document}
\title{Macroscopic Resonant Tunneling through
Andreev Interferometers}
\author{M.C.~Goorden}
\affiliation{D\'epartement de Physique Th\'eorique,
Universit\'e de Gen\`eve, CH-1211 Gen\`eve 4, Switzerland}
\author{Ph.~Jacquod}
\affiliation{Physics Department,
   University of Arizona, 1118 E. 4$^{\rm th}$ Street, Tucson, AZ 85721, USA}
\author{J.~Weiss}
\affiliation{Physics Department,
   University of Arizona, 1118 E. 4$^{\rm th}$ Street, Tucson, AZ 85721, USA}
\date{\today}
\begin{abstract}
We investigate the conductance through and the spectrum of ballistic
chaotic 
quantum dots attached to two $s$-wave superconductors, as a function of the
phase difference $\phi$ between the two order parameters. 
A combination
of analytical techniques -- random matrix theory, 
Nazarov's circuit theory and the
trajectory-based semiclassical theory -- allows us to explore the 
quantum-to-classical crossover in detail. When the superconductors
are not phase-biased, $\phi=0$, we recover known results that the
spectrum of the quantum dot exhibits an excitation gap, while the
conductance across two normal leads carrying $N_{\rm N}$ channels 
and connected to the dot
via tunnel contacts of transparency $\Gamma_{\rm N}$
is $\propto \Gamma_{\rm N}^2 N_{\rm N}$. 
In contrast, when $\phi=\pi$, the excitation gap closes and the
conductance becomes $G \propto \Gamma_{\rm N} N_{\rm N}$ in the 
universal regime.
For $\Gamma_{\rm N} \ll 1$, we observe 
an order-of-magnitude enhancement of the conductance towards
$G \propto N_{\rm N}$ in the
short-wavelength limit.
We relate this enhancement to resonant tunneling through 
a macroscopic number of levels
close to the Fermi energy. Our predictions are corroborated by numerical
simulations.
\end{abstract}
\pacs{74.45.+c, 73.23.-b, 74.78.Na, 05.45.Mt}
\maketitle
\section{Introduction}

\begin{figure}[t]
\begin{center}
\includegraphics[width=3cm]{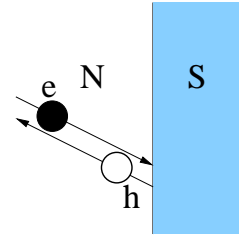}
\end{center}
\caption{ Andreev reflection by a superconductor (S). An incoming electron (e) in a
normal metal (N) near the Fermi energy $E_{\rm F}$ is reflected as a hole (h) with
opposite velocity.}
\label{andreev}
\end{figure}

Andreev reflection plays a central role in the description of
hybrid 
nanostructures with both superconducting and normal components~\cite{And64}. 
The process is illustrated in Fig.\ \ref{andreev}. When a negatively charged
electron in the normal metal hits the interface with a superconductor,
it is retroreflected into a positively charged hole. The superconductor acts as
a charge-inverting mirror, which leads to a variety of intriguing
effects in transport and spectroscopy of hybrid nanostructures~\cite{Imry}.
Here, we show how Andreev reflection resonantly
enhances the tunneling conductance through a quantum dot
by orders of magnitude, when the dot is coupled
to two superconductors with phase difference $\phi=\pi$ between their
order parameters. The effect
quickly disappears as $\phi$ is tuned away from resonance and
might have applications in low-temperature
current switching devices and magnetic flux ``transistors'', where $\phi$
would be controlled by an external magnetic flux. 

In a chaotic 
Andreev billiard, an impurity-free quantum dot coupled to one or more
superconductors, Andreev reflection renders all orbits periodic near the Fermi
level~\cite{Kos95}. Periodic orbits consist of an electron and a hole 
retracing each other's path, and therefore touching the superconductor
at both ends. This fact 
suggests that the density of states of an Andreev billiard
can be determined by
Bohr-Sommerfeld quantization. 
For a single superconducting lead, this results in
a density of states which is exponentially suppressed for
low energies, on the scale of the 
Thouless energy $E_{\rm T,S}=\hbar/2\tau_{\rm D,S}$,
where $\tau_{\rm D,S}$ is the average time between Andreev 
reflections~\cite{Mel96,Sch99}. Simultaneously,
one expects random matrix theory (RMT) to be valid
for chaotic quantum dots~\cite{BGS}.
RMT however leads to a different prediction, that a hard 
gap in the density of states opens up at
an energy $E_0\simeq 0.6 E_{\rm T,S}$\cite{Mel96}.
It was realized in Ref.\ \cite{Lod98} that the discrepancy between the
theories was not a short-coming in one of them but indicates 
that they have different regimes of validity. 
The crossover between the two regimes is determined by the Ehrenfest time,
which is the time scale for an initially minimal wave packet to spread over
the system. For a closed chaotic billiard of linear size $L_{\rm c}$ 
and Lyapunov exponent $\lambda$, 
the Ehrenfest time reads
$\tau_{\rm E} = \lambda^{-1} \ln [k_{\rm F} L_{\rm c}]$, with 
the Fermi wavenumber $k_{\rm F}$ of the quasiparticles~\cite{Sch05}.

RMT is valid in the {\it universal
regime} of vanishing Ehrenfest time, $\tau_{\rm E}\ll\tau_{\rm D,S}$,
while Bohr-Sommerfeld quantization gives a good description
in the opposite {\it semiclassical regime}, $\tau_{\rm E}\gg \tau_{\rm D,S}$.
Previous investigations of the crossover from 
the RMT regime to the semiclassical regime in 
Andreev billiards have considered the
density of states in the presence of a single $s$-wave superconductor, 
focusing on
the reduction of the excitation gap in the semiclassical regime, as well as
its fluctuations~\cite{Sch05,Ada02,Vav03,Sil03,Jac03,Goo03,Kor04,Goo05,Silv06,
Mel97,Zho98,Tar01}. 
The Ehrenfest time effects are usually small
and it takes a large computational effort to extract them numerically.

In this article we investigate
transport via normal metallic leads through an
Andreev billiard connected to two $\phi$-biased superconductors.
The system we consider is shown in Fig.\ \ref{cavity}.
A chaotic quantum
dot with mean level spacing $\delta$ is coupled to two superconductors. The
superconducting lead $1$ ($2$) has order parameter $\Delta_0e^{i\phi_1}$ 
($\Delta_0e^{i\phi_2}$), and we take 
$\Delta_0 \in {\cal R}$. One way to control
the phase difference $\phi\equiv \phi_1-\phi_2$ 
is to have them form a loop, which one then threads with a magnetic 
flux $\Phi$ -- this is sketched in Fig.\ \ref{cavity}. Alternatively,
the phase difference can be generated by a supercurrent.
In this case $\phi\equiv \phi_1-\phi_2=2 \pi \Phi/\Phi_0$, in terms of the
flux quantum $\Phi_0 = h/2e$. 
Since any global phase can be gauged out, we set $\phi_1=\phi/2$ and 
$\phi_2=-\phi/2$.
We assume that each
superconducting lead has $N_{\rm S}$ channels connected to the
quantum dot via tunnel contacts with tunnel probability
$\Gamma_{\rm S}$. The average time 
$\tau_{\rm D,S}=\hbar/2 E_{\rm T,S}$ 
between Andreev reflections 
is associated with the superconducting Thouless energy
$E_{\rm T,S}=2N_{\rm S}\Gamma_{\rm S}\delta/4\pi$.
Quasiparticle excitations have energy $E$ measured from the Fermi
energy $E_{\rm F}$, with $|E|\ll\Delta_0$. Thus the quasiparticles 
cannot penetrate into the superconductors. We
also assume $\Delta_0\ll E_{\rm F}$, in which case Andreev reflection 
perfectly retroreflects electrons into holes and vice versa,
with a phase shift $-\pi/2\pm \phi_i$. At $\phi=0$,
this {\it Andreev phase} shift
is solely due to the penetration of the wavefunction into the superconductor.
The additional 
phase shift $\pm \phi_i$ is due to the global phase of the superconductor
at contact $i$, where the minus
sign is for reflection from electron to hole, and the plus sign for reflection
from hole to electron.
We will refer to the 
quantum dot with two superconducting leads (but no
normal lead) as the closed Andreev billiard.
For the transport set-up we connect the quantum dot to two external
normal leads through a tunnel contact with tunnel
probability $\Gamma_{\rm N}$.
The current is measured between the right (R) and left (L) lead,
which both carry $N_{\rm N}$ transport channels.

The combination of Andreev 
and Ehrenfest physics in such an {\it Andreev interferometer} gives rise to
an order-of-magnitude enhancement of the tunneling conductance when $\phi=\pi$.
In the tunneling regime
$\Gamma_{\rm N} \rightarrow 0$, the conductance is strongly affected by
the density of states of the closed Andreev billiard.
In the absence of superconductivity ($\Gamma_{\rm S}=0$) and for 
$\Gamma_{\rm N} N_{\rm N} \gtrsim 1$ (we restrict ourselves to that
regime to avoid Coulomb blockade effects), the 
conductance is just the classical series conductance
$
G=G_0\Gamma_{\rm N} N_{\rm N}/2+{\cal O}(1), 
$
where $G_0=2e^2/h$~\cite{Bee97}.
In contrast, we will see below that in the limit 
$\tau_{\rm E}/\tau_{\rm D} \sim 1$, and for well coupled
superconductors, $\Gamma_{\rm S} \simeq 1$, 
the conductance at $\phi=\pi$ becomes
$G \propto G_0 N_{\rm N}$.
We identify the mechanism behind this enhancement as
{\it macroscopic resonant tunneling} through a large number
of low-energy quasi-degenerate Andreev levels very close to the
Fermi energy. Indeed, for $\phi=\pi$, all periodic
trajectories touching both superconductors, 
irrespective of their lengths, contribute to the density of
states at the Fermi energy in the semiclassical regime. Such a large effect is
already clearly visible for moderate Ehrenfest time, 
in contrast to previous 
works which found rather small manifestations of Ehrenfest physics
for larger $\tau_{\rm E}/\tau_{\rm D}$
~\cite{Ada02,Vav03,Sil03,Jac03,Goo03,Kor04,Goo05,Silv06}.
While similar behaviors were reported 
for cavities without transport~\cite{Kad95} or internal~\cite{Blom} 
mode-mixing, 
this effect was not found in earlier analytical
works on transport through chaotic Andreev 
interferometers~\cite{Spi82,Zai94,Bee95,Kad99}.

{\begin{figure}[t]
\begin{center}
\psfrag{Gn}{$\Gamma_{\rm N}, N_{\rm N}$}
\psfrag{Gs}{$\Gamma_{\rm S}, N_{\rm S}$}
\psfrag{R} {$R$}
\psfrag{L} {$L$}
\psfrag{P} {$\Phi$}
\psfrag{phi1} {$-\phi/2$}
\psfrag{phi2} {$\phi/2$}
\psfrag{1} {$S_1$}
\psfrag{2} {$S_2$}
\includegraphics[width=8cm]{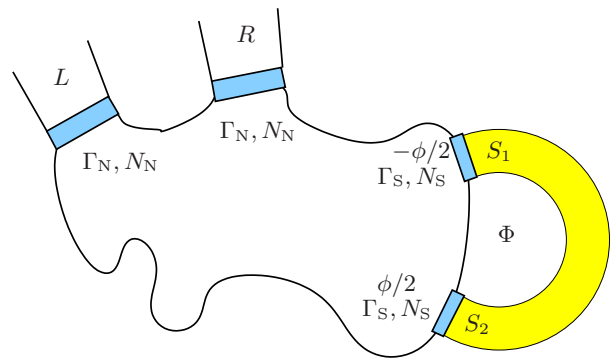}
\end{center}
\caption{ Quantum dot doubly coupled to a superconducting ring. Both 
contact to the superconductor have $N_{\rm S}$
channels and tunneling probability $\Gamma_{\rm S}$. 
The flux threading the ring causes a phase
difference $\phi=2\pi\Phi/\Phi_0$ between the two superconducting contacts.
Transport through the structure is investigated between two external
normal leads ($L$ and $R$), each carrying $N_{\rm N}$ transport channels, 
with tunnel contact of tunnel probability $\Gamma_{\rm N}$.}
\label{cavity}
\end{figure}

The paper is organized as follows. In section II 
we calculate the
density of states of the closed Andreev billiard. We repeat previous 
results in the
RMT limit and calculate the Bohr-Sommerfeld density of states as a function
of $\phi$. We show that due to a special resonance condition, the density of
states for phase differences close to $\pi$ has a large peak around the
Fermi energy.  
We proceed with a calculation of the conductance in Section \ref{sectionG}. 
In the RMT limit
we use Nazarov's circuit theory~\cite{Naz94,Arg96}. For weakly coupled normal 
leads, we find
that the conductance is small, $\propto \Gamma_{\rm N}^2 N_{\rm N}$, 
for $\phi\ll \pi$, while it reaches the
classical series conductance for $\phi=\pi$, because of 
the disappearance of the gap in the density of states. In the 
semiclassical limit we
employ a semiclassical trajectory-based 
method~\cite{Jac06,Semicl,Ada03,Bro06a,Whi07}. We find that the large peak
in the density of states at the Fermi level 
has its counterpart in the conductance, which
is a factor $1/\Gamma_{\rm N} \gg 1$ larger than in the RMT limit.
In section IV, numerical simulations confirm the validity of our 
analytical results.

\section{Spectrum of Andreev billiards}
\label{sectionrho}
\subsection{Random matrix theory}

The density of states in the universal regime has been calculated in Refs.\
\cite{Mel96,Mel97} using RMT. We repeat the main steps of 
the calculation for $\Gamma_{\rm S}=1$, and refer the reader 
to Refs.~\cite{Mel96,Fra96} for the calculation at arbitrary $\Gamma_{\rm S}$. 
At low excitation energy $E \ll \Delta_0$, 
the spectrum of
Andreev billiards is most conveniently  obtained from the 
low-energy effective
Hamiltonian~\cite{Fra96}
\begin{subequations}\label{eff-hamilton}
\begin{eqnarray}
{\cal H} &=& \left(
\begin{array}{cc}
H & -\pi W W^T \\
-\pi W W^T & -H^*
\end{array}
\right) \, , \\
W_{mn} &=& \delta_{mn} \left(\frac{M \delta}{\pi^2}\right)^{1/2}, \\
&& m=1,2,\ldots M, \;\;\;\;\, n=1,2,\ldots 2N_{\rm S}, \nonumber
\end{eqnarray}
\end{subequations}
which is constructed from the quantum dot's one-quasiparticle $M \times M$
Hamiltonian matrix $H$
and the projection matrix $W W^T$ giving the coupling between the quantum
dot and the superconductors. 
We define a $2 \times 2$ Green function $G(z)$ from the
matrix Green function
${\cal G}(z) = \langle (z-{\cal H})^{-1} \rangle$ 
\begin{eqnarray}
G = \left(
\begin{array}{cc}
G_{11} & G_{12} \\
G_{21} & G_{22}
\end{array}
\right) = 
\frac{\delta}{\pi} \left(
\begin{array}{cc}
{\rm Tr} {\cal G}_{11} & {\rm Tr} {\cal G}_{12} \\
{\rm Tr} {\cal G}_{21} & {\rm Tr} {\cal G}_{22}
\end{array}
\right),
\end{eqnarray}
from which the density of states reads
\begin{equation}
\label{rhoRMT}
\rho_{\rm RMT}(E)=-\frac{2}{\delta}{\rm Im}[G_{11}(E+i0^+)].
\end{equation}
Here, $G_{11}$ is self-consistently determined by the 
equations
\begin{subequations}\label{Greenfunction}
\begin{align}
\label{Greenfunction1}
G_{11}(z)=&-\frac{2\pi z}{N_{\rm S}\delta}\left(G_{12}^2(z)+\frac{G_{12}(z)}{\cos{(\phi/2)}}\right),\\
\label{Greenfunction2}
G_{12}^2(z)=&1+G_{11}^2(z).
\end{align}
\end{subequations}
For $\phi=0$, the density of states has a gap at an energy
$E_0\approx 0.6E_{\rm T,S}$. The energy of the gap is reduced when $\phi$ is
increased, closing at a value $\phi=\pi$, where one finds a flat density of
states $\rho(E)=2/\delta$. 

The theory of Refs.\ \cite{Mel96,Mel97} is only valid
for energies larger than the level spacing $\delta$. For a phase difference
$\phi=\pi$, the system
belongs to the symmetry class CI of Ref.\ \cite{Zir97}, and the 
density of states reads
\begin{eqnarray}
\label{rhozir}
\rho(E)=\frac{\pi}{\delta}\int_0^{2\pi E/\delta}dt J_0(t)J_1(t)/t, \\
\nonumber
\end{eqnarray}
with Bessel functions $J_0$ and $J_1$.
This prediction retains its validity down to low excitations $E < \delta$.
For energies $E\gg\delta$ (but still $E \ll \Delta_0$),
 Eq.\ (\ref{rhozir}) reduces to the mean field
theory and gives $\rho(E)=2/\delta$, while for $E\lesssim\delta$ it predicts 
a linearly vanishing density of states $\rho(E)=\pi^2 E/\delta$.

\subsection{Semiclassical regime}
The retroreflection at the superconductor makes all classical
trajectories near the
Fermi level periodic \cite{Kos95}. The elements of a periodic trajectory are 
an electron and a hole segment
retracing each other, and touching a superconducting contact
at both ends. The phase accumulated along a periodic orbit of period $T$
in one period consists of (i) the phase $2ET/\hbar$ acquired during 
the motion in the normal region and (ii) 
the phase shift acquired at the two Andreev
reflections. This shift equals $-\pi$ if the two reflections
take place at the
same superconducting lead and $-\pi\pm\phi$ if they
take place at different superconductors.
Summing the different contributions and requiring that the phase
accumulated in one period is a multiple of $2\pi$ leads to the 
Bohr-Sommerfeld quantization conditions 
\begin{subequations}
\begin{eqnarray}
\label{cond1}
E_n&=&\pi\hbar(n+ 1 / 2) \, \Big/ \, T,\\
\label{cond2}
E_n&=&\pi\hbar(n+1 / 2+ \phi / 2\pi) \, \Big/ \, T,\\
\label{cond3}
E_n&=&\pi\hbar(n+ 1 / 2 -\phi / 2\pi ) \, \Big/ \, T.
\end{eqnarray}
\end{subequations}
Eq.~(\ref{cond1}) applies to trajectories touching
the same superconductor twice, while Eqs.~(\ref{cond2}) and (\ref{cond3}) 
to trajectories touching both superconductors, depending on whether 
the Andreev reflection from electron to hole takes 
place at the superconductor with phase $\phi/2$ 
[Eq.~(\ref{cond2})] or $-\phi/2$ [Eq.~(\ref{cond3})]. Assuming ergodicity,
the corresponding trajectories have a weight $1/2$, $1/4$, $1/4$, 
leading to a mean density of states (see also 
Ref.~\cite{Ihra01} for a single superconducting contact)
\begin{widetext}
\begin{equation}
\label{rhoBS}
\rho_{\rm BS}(E)=N_{\rm S}\int_0^\infty
dTP(T) \sum_{n} 
\left[
\delta\left(E-[n+\frac{1}{2}]\frac{\pi\hbar}{T}\right)+
\frac{1}{2}\delta\left(E-[n+\frac{1}{2}+\frac{\phi}{2\pi}]\frac{\pi\hbar}{T}\right)+\frac{1}{2}\delta\left(E-[n+\frac{1}{2}-\frac{\phi}{2\pi}]\frac{\pi\hbar}{T}\right)\right].
\end{equation}
Here $P(T)$ is the classical distribution of return times to a
superconducting contact. Inserting 
the chaotic distribution 
$P(T)=\exp{(-T/\tau_{\rm D,S})} \big/ \tau_{\rm D,S}$ into
Eq.\ (\ref{rhoBS}) gives
\begin{align}
\label{rhoBScalc}
\rho_{\rm BS}(u)=&\frac{2\pi}{\delta}\frac{\cosh{\left(\frac{\phi}{2u}\right)}}{\sinh{\left(\frac{\pi}{u}\right)}u^2}
\left(\pi\cosh{\left(\frac{\phi}{2u}\right)}\coth{\left(\frac{\pi}{u}\right)}-\phi\sinh{\left(\frac{\phi}{2u}\right)}\right),
\end{align}
with $u=E/E_{\rm T,S}$.
For
$\phi=0$ Eq.\ (\ref{rhoBScalc}) reduces to the Bohr-Sommerfeld approximation of Refs.\
\cite{Mel96, Sch99}
and the density of states is exponentially suppressed at low
energies. 
\end{widetext}

\subsection{Comparison of the two regimes}
\begin{figure}
\begin{center}
\includegraphics[width=8.5cm]{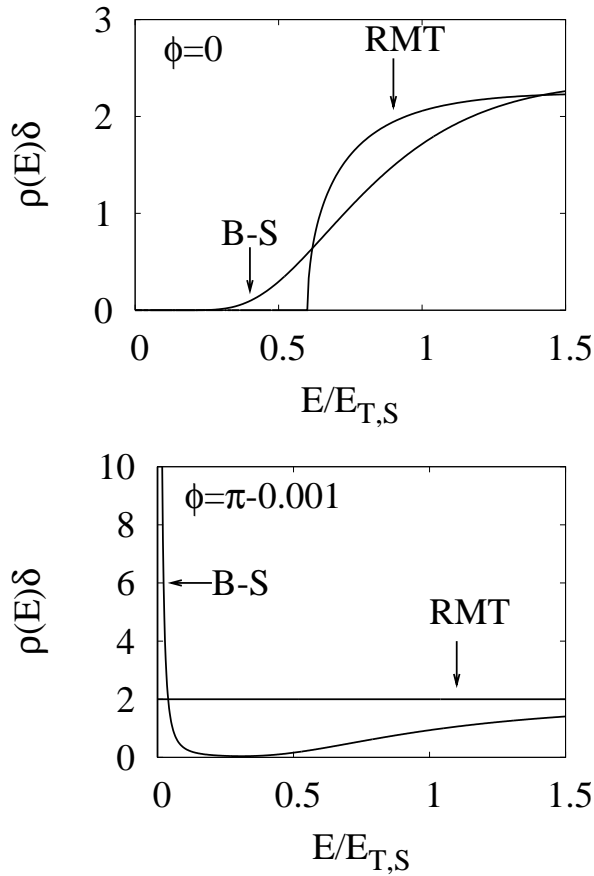}
\end{center}
\caption{Comparison of the density of states calculated from
Bohr-Sommerfeld quantization
and random matrix theory (RMT). The top panel is for 
a phase difference $\phi=0$,
and the bottom panel for $\phi\approx\pi$.}
\label{rhoan}
\end{figure}

The low energy approximation of Eq.\ (\ref{rhoBScalc}) reads
\begin{equation}
\label{rhoBSlowE}
\lim_{u\rightarrow 0} \rho_{\rm BS}(u)=\frac{2\pi}{\delta}\frac{e^{-(\pi-\phi)/u}(\pi-\phi)}{2u^2}.
\end{equation}
For $\pi-\phi\ll 1$ this gives a sharply peaked function with a maximum value of $\rho_{BS}=\frac{4\pi}{e^2(\pi-\phi)}$ at energy $u=\frac{\pi-\phi}{2}$. In the limit $\phi\rightarrow\pi$ we find
\begin{equation}
\label{rhoBSpi}
\lim_{u\rightarrow 0}\lim_{\phi\rightarrow\pi} \rho_{\rm BS}(u)=\frac{2\pi}{\delta}\delta(u).
\end{equation}
Eq.~(\ref{rhoBSpi}) predicts a
number ${\cal O}[N_{\rm S}]$ of levels around $E=0$.
The peculiar behavior at $\phi=\pi$ is due to the fact that 
the two quantization conditions (\ref{cond2}) and (\ref{cond3}) 
have solutions $E = 0$ simultaneously for {\em all}
trajectories, irrespective of their period.

Fig.\ \ref{rhoan} illustrates the predictions of the previous two paragraphs.
In the RMT limit we
combine the solution of Eqs.\ (\ref{Greenfunction1}) and (\ref{Greenfunction2})
with Eq.\ (\ref{rhoRMT}), while in the semiclassical limit we use
Eq.\ (\ref{rhoBS}). For $\phi=0$ (top panel), the density of states in the RMT and semiclassical limit are very similar.
Both densities of states are suppressed at the Fermi energy, 
while they are restored to a value of $2/\delta$ at an energy scale set 
by $E_{\rm T,S}$. The main difference is in the way that $\rho$ is suppressed: a 
hard gap in the RMT limit vs. an exponential suppression for the semiclassical 
density of states. 

In contrast, when $\phi \approx \pi$, the
difference between RMT and Bohr-Sommerfeld predictions is huge. 
While in the RMT limit
the density of states is flat 
[up to the minigap of order $\delta$ predicted by Eq.(\ref{rhozir}) which is
not resolved in Fig.~\ref{rhoan}], 
it shows a large peak around
the Fermi energy in the semiclassical limit.  
For $\phi=\pi$ the peak is described by a Dirac $\delta$-function (cf. Eq.\
(\ref{rhoBSpi})), which is why we show a
density of states at a phase difference slightly different from $\pi$
in Fig.\ \ref{rhoan}.

\section{Transport}
\label{sectionG}
\subsection{Random matrix theory}
\label{sectionGrmt}
\begin{figure}
\begin{center}
\psfrag{N} {$N$}
\psfrag{S} {$S$}
\psfrag{1} {$1$}
\psfrag{2} {$2$}
\psfrag{3} {$3$}
\psfrag{4} {$4$}
\psfrag{5} {$5$}
\psfrag{Nn} {$N_{\rm N},\Gamma_{\rm N}$}
\psfrag{Ns} {$N_{\rm S},\Gamma_{\rm S}$}
\includegraphics[width=2.5cm]{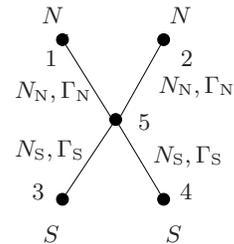}
\end{center}
\caption{Circuit for our system. We have two normal electrodes ($1,2$), two
  superconducting electrodes ($3,4$) and one internal junction, the chaotic
  cavity ($5$). The
  superconducting  (normal) leads have $N_{\rm S}$ ($N_{\rm N}$) 
channels with transmission probability $\Gamma_{\rm S}$ ($\Gamma_{\rm N}$). }
\label{circuit}
\end{figure}
We use Nazarov's circuit
theory of Andreev conductance \cite{Naz94} to calculate
the two-terminal conductance through the system shown in Fig.~\ref{cavity}  
(see also Ref.\ \cite{Arg96}). 
The theory describes hybrid nanostructures in the same language as electrical 
circuits: i.e. their structure is reduced 
to internal junctions, normal/ superconducting external junctions (electrodes)
and leads connecting the different elements. The leads are characterized
by a set of transmission eigenvalues.
The conductance between different electrodes is determined by a simple set of
algebraic equations.
As the energy dependence of the scattering matrix is neglected,
the theory is valid for small temperatures and voltages. 
The superconductors have 
the same voltage. The results are valid to leading order in
$\Gamma_{\rm S} N_{\rm S}$, $\Gamma_{\rm N} N_{\rm N}$.

Our system consists of two
normal external junctions (denoted by indices $1,2$), two
superconducting external junctions $3,4$ and one internal junction, the
chaotic cavity $5$. This is illustrated in Fig.\ \ref{circuit}.
The calculation of the conductance starts with the assignment of a complex amplitude
$f_k$ to each junction. The amplitudes are somewhat analogous to voltages: if
a junction $n$ is connected to a junction $m$ by a lead with a large
conductance, then $f_n \rightarrow f_m$. 
The physical meaning of the amplitudes is the following: if electrons are
injected into the $k^{\rm th}$ junction via an additional weakly coupled normal
electrode, a beam of holes with intensity $|f_k|^2$ times that of the electron
beam is retro-reflected into that probe.
For the external junctions the value of the amplitudes is 
prescribed by the theory: 
for the normal external junctions $f_1$,$f_2=0$ and 
for the superconducting junctions $f_3=\exp{(-i\phi/2)}$, 
$f_4=\exp{(i \phi/2)}$. 
The amplitude $f_5$ can be found by the spectral current
conservation rule~\cite{Arg96}. One has
$f_5=\tan(\theta_5/2)$, with $\theta_5$ to be 
determined by 
\begin{eqnarray}
\label{theta5}
\frac{\Gamma_{\rm S} N_{\rm S}}{\Gamma_{\rm N} N_{\rm N}}\cos{\frac{\phi}{2}}\cos{\theta_5}(1+\cos{\theta_5})\left(1+(1-\Gamma_{\rm N})\tan^2{\frac{\theta_5}{2}}\right) \nonumber\\=\sin\theta_5(2-\Gamma_{\rm S}+\Gamma_{\rm S}\sin{\theta_5}\cos{\frac{\phi}{2}}). \qquad
\end{eqnarray}
Once the complex amplitudes are known, the RMT-averaged 
conductance between junctions $1$ and $2$ is given by
\begin{equation}
\label{G12}
\langle G \rangle/G_0 =\frac{\Gamma_{\rm N} N_{\rm N}}{2}\frac{\cos{\theta_5}+\Gamma_{\rm N}\sin^2[\theta_5/2]}{(1-\Gamma_{\rm N}\sin^2[\theta_5/2])^2}.
\end{equation}
The self-consistency equation (\ref{theta5}) for $\theta_5$ combined with
Eq.\ (\ref{G12}) determines the conductance.
For ballistic leads ($\Gamma_{\rm N}=\Gamma_{\rm S}=1$) we reproduce
the results of Ref.\ \cite{Bee95}.

Of most interest to us is the regime $\Gamma_{\rm N} \ll 1$
and $\Gamma_{\rm S}=1$,
where we expect a strong connection between conductance and density of states.
In the limit 
$\delta \phi=\pi-\phi\gg\Gamma_{\rm N}$,
we use an expansion in terms of $\Gamma_{\rm N}$, while in the
opposite limit, we first expand in the small parameter $\delta \phi$ . 
From Eqs.\
(\ref{theta5}) and (\ref{G12})
we find, to leading order in $\Gamma_{\rm N}$,
\begin{align}
\label{G12ex}
\langle G \rangle/G_0 =
\left\{\begin{array}{cc}
\Gamma_{\rm N}^2 N_{\rm N} \left(1+\frac{N_{\rm N}}{N_{\rm
    S}}\frac{1+\cos{\frac{\phi}{2}}}{\cos{\frac{\phi}{2}}}\right) \Big/
4,
&\mbox{if}\hspace{0.1cm} \delta\phi\gg\Gamma_{\rm N},\\
\\
\Gamma_{\rm N} N_{\rm N} / 2+{\cal O}(\delta\phi^2),&\mbox{if}\hspace{0.1cm} \delta \phi\ll\Gamma_{\rm N}.
\end{array}\right.
\end{align}
The main feature of
Eq.\ (\ref{G12ex}) is
that, for $\phi=\pi$, one has the same conductance as a
two-terminal quantum dot without superconductor. This result
agrees with the density of states where the 
influence of the superconductors also disappears and the density of states
becomes flat (bottom panel of Fig.\ \ref{rhoan}). The minigap of magnitude
$\delta$ is hardly resolved, due to the coupling to external leads and the 
associated broadening of levels.
On the other hand, for $\phi\ne\pi$ the density of states is strongly suppressed at the Fermi level (top panel of Fig.\ \ref{rhoan}) and we expect a conductance which is reduced due to the presence of the superconductors. Eq.\
(\ref{G12ex}) does indeed predict a conductance of order $\Gamma_{\rm N}^2$. 

From Eq.\ (\ref{G12ex}) it follows that the crossover between the two limits 
occurs when $\delta\phi=\Gamma_{\rm N} N_{\rm N}/N_{\rm S}$.
We now connect this to the density of states of the Andreev billiard.
Taylor expanding Eqs.~(\ref{rhoRMT}), (\ref{Greenfunction1}) and (\ref{Greenfunction2}) in $\delta \phi$ gives a gap in the density of states at an energy 
\begin{equation}
\label{gap}
E_0=\delta \phi \, E_{\rm T,S}=\delta \phi \, \frac{N_{\rm S} \, \delta}{2\pi}.
\end{equation} 
The density of states of the closed cavity is broadened due to the
coupling to normal leads. The coupling to leads results in a
non-hermitian Hamiltonian
$H \rightarrow H-i\pi W_{\rm N}W_{\rm N}^\dag$, with $H$ the
Hamiltonian of the closed system and $W_{\rm N}$ an $M\times 2N_{\rm N}$ matrix describing 
the coupling to
the normal leads \cite{Bee97,Iida90}. The $2N_{\rm N}\times 2N_{\rm N}$ matrix $W_{\rm N}^\dag W_{\rm N}$ has 
eigenvalues 
$w= \Gamma_{\rm N} M \delta / 4\pi^2 +{\cal O}(\Gamma_{\rm N}^2)$. 
For the broadening we need the eigenvalues of the $M\times M$ matrix
$W_{\rm N}W_{\rm N}^\dag$.  Assuming that all levels in the dot are coupled to
the leads in the same manner, we multiply $w$ by $2N_{\rm N}/M$ to estimate the
broadening as $\delta E=2 \pi w N_{\rm N} / M= \Gamma_{\rm N} N_{\rm N} \delta/ 2\pi $. 
Comparing this with Eq.\ (\ref{gap}) we find that $\delta E=E_0$ for
$\Gamma_{\rm N} N_{\rm N}=N_{\rm S}\delta\phi$. This gives the crossover 
energy scale for Eqs.\
(\ref{G12ex}). We conclude that the conductance is not affected by the
presence of the superconductors once the gap in the broadened density of 
states closes.

We finally comment on the effect of a finite tunnel barrier 
with $\Gamma_{\rm S} < 1$
between the superconductor and normal metal. 
For $\Gamma_{\rm N} \ll 1$, Eq.~(\ref{G12}) becomes
\begin{align}
\label{G12exGammas}
\langle G \rangle/G_0=
\left\{\begin{array}{cc}
\Gamma_{\rm N}^2 N_{\rm N} \left(1+\frac{N_{\rm N}}{N_{\rm S,\rm eff}}\frac{1+\cos{\frac{\phi}{2}}}{\cos{\frac{\phi}{2}}}\right) \Big/4, 
&\mbox{if}\hspace{0.1cm} \delta\phi\gg\Gamma_{\rm N},\\
\\
\Gamma_{\rm N}  N_{\rm N}/ 2+{\cal O}(\delta\phi^2), \; &\mbox{if}\hspace{0.1cm} \delta\phi\ll\Gamma_{\rm N}.
\end{array}\right.
\end{align}
Here, $N_{\rm S,\rm eff}=c \, \Gamma_{\rm S} N_{\rm S}$ where
$c\in (\frac{1}{2},1)$ is a numerical factor depending only weakly on $\phi$
and $\Gamma_{\rm S}$. A finite
$\Gamma_{\rm S}$ thus reduces the number of
effective superconducting channels to $N_{\rm S,{\rm eff}}$.

\subsection{Semiclassical regime}

The starting point of our semiclassical treatment is
the two-terminal conductance 
through a metallic system in contact with superconductors
(see Fig.~\ref{fig:device})~\cite{Lam93},
\begin{equation}\label{eq:full-lambert}
G /G_0 = T_{RL}^{ee}+T_{RL}^{he}+
2 \frac{T_{LL}^{he} T_{RR}^{he}
-T_{LR}^{he} T_{RL}^{he}}{T_{LL}^{he}+T_{RR}^{he}+T_{LR}^{he}+T_{RL}^{he}}.
\end{equation}
We used the transmission probability for a
quasi-particle of type $\alpha=e,h$ from the normal lead $i=L,R$ to a 
quasi-particle of type $\beta=e,h$ into the normal lead $j=L,R$,
\begin{equation}\label{eq:ampl_probs}
T_{ji}^{\beta \alpha}= \sum_{m \in i;n \in j} 
|\tau^{\beta \alpha}_{n,j;m,i}|^2.
\end{equation}
Here, $\tau^{\beta \alpha}_{n,j;m,i}$ gives
the transmission amplitude from 
channel $m$ in lead $i$ to channel $n$ in lead $j$. Since 
we consider a symmetric configuration, where both normal leads carry the same
number $N_{\rm N}$ of channels, all of them coupled to the cavity with the
same tunnel probability $\Gamma_{\rm N}$, Eq.~(\ref{eq:full-lambert})
simplifies to
\begin{equation}\label{eq:sym_lambert}
\langle G \rangle /G_0 = \langle T_{RL}^{ee} \rangle + \langle T_{LL}^{he} 
\rangle.
\end{equation}
Eq.~(\ref{eq:sym_lambert})
gives the ensemble-averaged conductance to leading
order in $\Gamma_{\rm N} \, N_{\rm N} \gg 1$. 

\begin{figure}[ht]
\begin{center}
\psfrag{Gamma}{\Large $\Gamma_{\rm N}, N_{\rm N}$}
\psfrag{gs1}{\Large $\gamma_{s1}$}
\psfrag{gs2}{\Large $\gamma_{s2}$}
\psfrag{gs3}{\Large $\gamma_{s3}$}
\psfrag{gn}{\Large $\gamma_{\rm n}$}
\psfrag{R} {\Large $R$}
\psfrag{L} {\Large $L$}
\psfrag{S1} {\Large $S_1$}
\psfrag{S2} {\Large $S_2$}
\psfrag{D1}{\Large $\Delta e^{-i \phi/2}$}
\psfrag{D2}{\Large $\Delta e^{ i \phi/2}$}
\psfrag{Gammas}{\Large $\Gamma_{\rm S}, N_{\rm S}$}
\includegraphics[width=9cm]{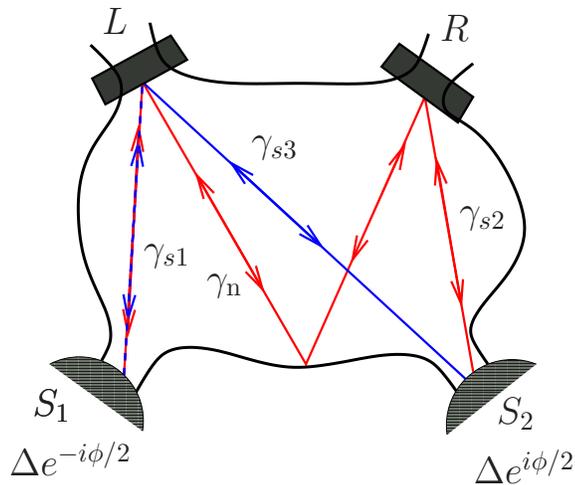}
\vspace{-1.1cm}
\end{center}
\caption{\label{fig:device} 
Schematic of the class I (blue) and class II (red) periodic orbits 
of the closed Andreev billiard, giving rise
to macroscopic resonant tunneling at $\phi=\pi$ when the
cavity is connected to external electrodes ($L$ and $R$).}
\end{figure}

To evaluate the resonant contributions to $\langle T_{RL}^{ee} \rangle$ and 
$\langle T_{LL}^{he} \rangle$,
we follow the semiclassical approach of Ref.~\cite{Jac06} (see also
Refs.~\cite{Semicl,Ada03,Bro06a,Whi07})
Semiclassically, the transmission amplitudes read,
\begin{eqnarray}
\tau^{\beta \alpha}_{n,j;m,i}
&=&
-(2\pi \rmi \hbar)^{-1/2}
\!\int_{\rm i} \! \! \rmd y_0 \int_{\rm j} \! \rmd y 
\sum_\gamma A_\gamma
\nonumber \\
& &\qquad \times \langle n|y\rangle  
\langle y_0| m \rangle 
\exp[\rmi S_\gamma /\hbar + \rmi \pi \mu_\gamma /2]
\, ,\qquad
\end{eqnarray}
where $|m\rangle$ is the transverse wavefunction
of the $m^{\rm th}$ lead mode. This expression 
sums over all  trajectories $\gamma$ (with classical action $S_{\gamma}$ 
and Maslov index $\mu_\gamma$) starting at $y_0$ on a cross-section of
the injection ($i$) lead and ending at $y$ on the
exit ($j$) lead, with even ($\alpha=\beta$) or odd ($\alpha \ne \beta$)
number of Andreev reflections. The transmission probabilities 
are then given by a double sum over trajectories.
After the semiclassical approximation that
$\sum_n\langle y'|n\rangle\langle n|y\rangle \simeq \delta (y'-y)$ 
\cite{Jac06} one has
\begin{eqnarray}\label{semicl-tr}
T_{ji}^{\beta \alpha} &=& \frac{1}{2 \pi \hbar} \int_i \rmd y_0 \int_j \rmd y_0'
\sum_{\gamma1,\gamma2} A_{\gamma1} A_{\gamma2}^* \exp[i \delta S/\hbar]
\, .\qquad
\end{eqnarray} 
This expression sums over all pairs of classical 
trajectories $\gamma1$ and $\gamma2$ with fixed endpoints 
$y_0$ and $y_0'$, which
convert an $\alpha$ quasiparticle into a $\beta$ quasiparticle. 
The phase 
$\delta S = S_{\gamma1}-S_{\gamma2}$
gives the difference in action phase accumulated along $\gamma1$ and $\gamma2$.
In the presence of tunnel barriers, the stability 
amplitudes $A_\gamma$ are given by~\cite{Whi07,Cou92}
\begin{equation}\label{eq:stability}
A_\gamma = B_\gamma \, t_i t_j \,
\prod_k [r_k]^{l_{\gamma}(k)},
\end{equation}
where $l_\gamma(k)$ gives the number of times that $\gamma$ is reflected
back into the system from the tunnel barrier at $k=L,R,S_1,$ and $S_2$, 
the transmission
and reflection amplitudes satisfy $|t_i|^2 = (1-|r_i|^2)=
\Gamma_{\rm N,S}$ (for $i=L, \, R, \, S_1,$ or $S_2$), and
$B_\gamma^2 = (d p_{y_0}/d y)_\gamma$ measures the rate of change of the 
initial
momentum $p_{y_0}$ as the exit position $y$ of $\gamma$ is changed,
for a fixed sequence of transmissions and reflections at the tunnel
barriers.

There are four types of trajectories to consider: \\[-2mm]

(i) paths that do not touch the superconductors,

(ii) paths that hit one of the superconductor once,

(iii) paths that hit the same superconductor twice,

(iv) paths that hit both superconductors. \\[-2mm]

\noindent Trajectories of type (iv) are flux-dependent. They are
depicted in Fig.~\ref{fig:device}, and it turns out that they
are the ones giving rise to macroscopic resonant tunneling.
Our semiclassical investigations focus on those paths. 

We use Eqs.~(\ref{semicl-tr}) and (\ref{eq:stability})
to evaluate the dominant semiclassical contributions
to Eq.~(\ref{eq:sym_lambert}). We subdivide the relevant trajectories
into class I trajectories, contributing to 
$\langle T_{LL}^{he} \rangle$ 
(blue trajectory on Fig.\ref{fig:device}), 
and class II trajectories, contributing to $\langle T_{RL}^{ee} \rangle$
(red trajectory on Fig.\ref{fig:device}).

Class I trajectories are made of the following sequence
\begin{equation}\label{eq:classI}
\gamma_I^{(p)} = \gamma^{(e)}_{s1}+\gamma^{(h)}_{s1}
+p \times \left[\gamma^{(h)}_{s3}+\gamma^{(e)}_{s3}+\gamma^{(e)}_{s1}+\gamma^{(h)}_{s1}
\right],
\end{equation}
where $s1$ and $s3$ can be interchanged, and $p=0,1,2,\ldots$
These trajectories undergo $2p+1$ Andreev reflections and $2p$ reflections at 
tunnel barriers~\cite{caveat}. They accumulate an action phase
\begin{equation}\label{eq:action_classI}
S_{\gamma,I} = p (-\pi +\phi + E \, t_{\ell \rm,I}) + 2 \, E \,
t_{\gamma_{s1}}-(\pi/2- \phi/2).
\end{equation}
One should substitute $\phi\rightarrow -\phi$ when interchanging
segments $s1$ and $s3$, though the relative sign 
between $\pi$ and $\phi$ does not affect the final result.
Here, $t_{\ell \rm,I}$ gives the duration of the loop [the 
sequence between bracket 
in Eq.~(\ref{eq:classI})] and $t_{\gamma_{s1}}$ is the duration of the 
segment $\gamma_{s_1}$ as shown on Fig.~\ref{fig:device}.
Class II trajectories split into two subclasses, defined by the following two
sequences
\begin{subequations}\label{eq:classII}
\begin{eqnarray}
\gamma_{IIa}^{(p)} &=& \gamma^{(e)}_{n} \\
&+& p \times 
\left[\gamma^{(e)}_{s2}+\gamma^{(h)}_{s2}+\gamma^{(h)}_{n}+\gamma^{(h)}_{s1}+
\gamma^{(e)}_{s1}+ \gamma^{(e)}_{n}\right] \, , \nonumber \\
\gamma_{IIb}^{(p)} &=& \gamma^{(e)}_{s1}+\gamma^{(h)}_{s1}
+\gamma^{(h)}_{n}+\gamma^{(h)}_{s2}+\gamma^{(e)}_{s2} \\
&+&p \times 
\left[\gamma^{(e)}_{n}+\gamma^{(e)}_{s1}+\gamma^{(h)}_{s1}+\gamma^{(h)}_{n}
+\gamma^{(h)}_{s2}+\gamma^{(e)}_{s2}
\right] \, .\nonumber 
\end{eqnarray}
\end{subequations}
They undergo $2 p$ (IIA) and $2p+2$ (IIb) Andreev reflections, 
$4p$ (IIa) and $4p +2$ (IIb) reflections at tunnel barriers~\cite{caveat},
and accumulate action phases
\begin{subequations}\label{eq:action_classII}
\begin{eqnarray}
S_{\gamma,IIa} &=& p (-\pi - \phi + E \, t_{\ell \rm,IIa})+ (E_{\rm F}+E) \, 
t_{\gamma_{n}},
\label{eq:action_classIIa}\\
S_{\gamma,IIb} &=& p (-\pi +\phi + E\, t_{\ell \rm,IIb})- (E_{\rm F}-E) \,
t_{\gamma_{n}} \nonumber \\
&&+2 E \, (t_{\gamma_{s1}}+t_{\gamma_{s2}})
-(\pi - \phi).
\label{eq:action_classIIb}
\end{eqnarray}
\end{subequations}
Here $t_{\ell \rm,IIa}$ and $t_{\ell \rm,IIb}$ give the duration of the Andreev loops
[the two sequences between bracket in Eqs.~(\ref{eq:classII})].
We already see that at $E=0$ and 
$\phi=\pi$, the phase difference accumulated by any two members
(with different $p$) of a given family vanishes, so that
all pair of trajectories within a given family 
resonate. There is however no resonance between
members of different families. 

In normal chaotic billiards, the stability $B_\gamma^2$ of
periodic orbits decreases exponentially with the number of times the
orbit is traveled~\cite{Haake-book}. The situation is fundamentally
different in presence of superconductivity, where Andreev reflections
refocus the dynamics. The stability of a trajectory is then only
given by the product of the stabilities
along the primitive segments, independent of $p$. 
We will take these primitive segments 
as $\gamma_{s1}$ and $\gamma_{s3}$ for class I,
$\gamma_{s1}$, $\gamma_n$ and $\gamma_{s2}$ for class II,
keeping in mind, however, that
the contributions arising from the trajectories of class I
and IIa with $p=0$ are more stable than those with $p \ge 1$, because they do 
not travel on $\gamma_{s3}$ (class I) nor on $\gamma_{s1}$ and $\gamma_{s2}$
(class IIa). Their contribution is thus underestimated in our approach,
and our final results, Eqs.~(\ref{eq:TLLhe}) and (\ref{eq:TRLee}),
underestimate the conductance by a 
subdominant correction.

This enhanced stability applies to trajectories whose 
Andreev loop is shorter than twice the Ehrenfest 
time, i.e. the time beyond which an initially narrow
wavepacket can no longer fit inside
a superconducting lead~\cite{Lod98,Jac03,Vav03}. 
The relative measure of trajectories of class I ($P_I$) and II ($P_{II}$) 
is thus
\begin{widetext}
\begin{subequations}\label{eq:probabilities}
\begin{eqnarray}
P_I (\tE/\tD) & = &
\tD^{-2} \int_0^{\tau_{\rm E}} \rmd t_{s1} 
\int_0^{\tau_{\rm E}-t_{s1}} \rmd t_{s3} \, \exp[-(t_{s1}+t_{s3})/\tD] = 
1-[1+\tau_{\rm E}/\tau_{\rm D,S}] \, \exp[-\tau_{\rm E}/\tau_{\rm D,S}], \\
P_{II} (\tE/\tD) & = &\tD^{-3} 
\int_0^{\tau_{\rm E}} \rmd t_{s1} 
\int_0^{\tau_{\rm E}-t_{s1}} \rmd t_{n} 
\int_0^{\tau_{\rm E}-t_{s1}-t_n} \rmd t_{s2} 
\, \exp[-(t_{s1}+t_n+t_{s3})/\tD] \nonumber \\
&=&1-[1+\tau_{\rm E}/\tau_{\rm D,S}+(\tau_{\rm E}/\tau_{\rm D,S})^2/2] \, \exp[-\tau_{\rm E}/\tau_{\rm D,S}] \, .
\end{eqnarray}
\end{subequations}

We are now ready to evaluate the dominant contributions to conductance
close to resonance. We focus on the low-temperature regime with
$E=0$. We start from Eq.~(\ref{semicl-tr}) and pair trajectories by class,
noting that for a given class, all 
trajectories have the same stability
but differ only by the number of Andreev reflections
and normal reflections at the contacts to the normal leads, 
as well as by the different
action phases they accumulate along
their Andreev loop. 
The sum over classes is then represented by a sum over 
primitive
trajectories, and we perform the substitution
\begin{eqnarray}\label{semicl-tr-spa}
\sum_{\gamma1,\gamma2} A_{\gamma1} A_{\gamma2}^* \; [ \dots ]_{\gamma1,\gamma2} \;
\longrightarrow  \; \Gamma_{\rm N}^2
\sum_{\gamma = {\rm primitive}} B_{\gamma}^2 \sum_{p,p' =0}^{\infty} 
(1-\Gamma_{\rm N})^{a(p+p')} \, \Gamma_{\rm S}^{p+p'+c} \; [ \dots ]_{\gamma,p,p'} 
\, .\\[-3mm]
\nonumber
\end{eqnarray}
\end{widetext} 
The exponents $a=1$, $b=0$ and $c=1$ 
for class I, $a=2$, $b=0$ and $c=0$ 
for class IIa and $a=2$, $b=2$ and $c=2$ 
for class IIb are determined by the number of Andreev and
normal reflections in Eqs.~(\ref{eq:classI}) and (\ref{eq:classII}).
Reflection phases do not appear because each time an electron is
reflected at a tunnel barrier, a hole is reflected later.
The argument in the double sum over $p$ and $p'$ in Eq.~(\ref{semicl-tr-spa})
has to be multiplied
by the $p$-$p'$--dependent Andreev phase.
The double sum over $p$ and $p'$ is easily resummed, 
\begin{eqnarray}\label{eq:resummation}
\sum_{p,p' =0}^{\infty} 
&&(1-\Gamma_{\rm n})^{a(p+p')+b} \; \Gamma_{\rm s}^{p+p'+c} \; e^{i(p-p')(\phi-\pi)} = \\
&& \frac{\Gamma_{\rm S}^c \; (1-\Gamma_{\rm N})^b}{1-2 \, \Gamma_{\rm S} \,
(1-\Gamma_{\rm N})^a
\cos[\pi-\phi]+\Gamma_{\rm S}^2 \, (1-\Gamma_{\rm N})^{2a}}.\nonumber
\end{eqnarray}
We next evaluate $\sum_{\gamma} B_{\gamma}^2$
by relating it to classical transmission probabilities~\cite{Jac06}. 
Because the sum runs over primitive trajectories whose stability
is given by their electronic (or hole) component alone, 
one considers that the cavity is
opened to four normal leads. 
The sum runs over all 
trajectories such as those sketched in Fig.~\ref{fig:classI}.
For class I, 
these trajectories are made up of two legs, both of them
starting at $y_0$ on the $L$ lead, and we write
$\sum B_\gamma^2 = \sum_{\gamma 1,\gamma 3} B_{\gamma1}^2
B_{\gamma 3}^2$. The leg $\gamma1$ makes a normal angle
$\theta_0$ with respect to the normal to a cross-section
of $L$ and ends up in lead $S_1$, 
while $\gamma 3$ makes a normal angle $\theta_0'=-\theta_0$ 
and goes to lead $S_2$. Both incidence
position and angle on $S_1$ and $S_2$ have to be integrated over.

We define $P({\bf Y},{\bf Y}_0;t)\de y\de \theta \de t$ as the 
classical probability to go from an initial position and momentum angle
${\bf Y}_0=(y_0,\theta_0)$ on a lead
to within $(\de y,\de \theta)$ of
${\bf Y}=(y,\theta)$ on another lead,
in a time within $\de t$ of $t$.
The sum over
all primitive trajectories of class I can then be rewritten as
\begin{eqnarray}
\sum_{\gamma, {\rm primitive}} \!
B_\gamma^2&=& \!\! p_{\rm F}
\int_0^{\tE} 
\! \rmd t_{s1} \int_0^{\tE-t_{s1}} 
\rmd t_{s3} \; \int_{-\pi/2}^{\pi/2} \rmd \theta_0 
\rmd \theta_0' \nonumber \\
&\times& \int_{S1} \rmd y_{s1} \int_{S2} \rmd y_{s3}
\int_{-\pi/2}^{\pi/2} \rmd \theta_{s1}
\rmd \theta_{s2} \nonumber \\
&\times&
\; P({\bf Y}_{s1},{\bf Y}_0;t_{s1}) \, P({\bf Y}_{s3},{\bf Y}_0;t_{s3}) 
\nonumber \\
&\times & \cos \theta_0 \, \cos \theta_0' \, \delta(\theta_0+\theta_0') \, \delta(y_0-y_0').
\label{eq:gamma-sum-to-Pintegral}
\end{eqnarray}
The factor $p_{\rm F} \cos \theta_0$ measures the injected current, and
since electrons are reflected (and not injected) at the $L$ tunnel barrier, 
one has an additional $\cos \theta_0'$ (instead of
$p_{\rm F} \cos \theta_0'$).
For a given cavity, $P$ is a sum of $\de$-functions over all 
possible classical trajectories. Instead we consider
the distribution averaged over a mesoscopic
ensemble of similar cavities or a small energy interval. This gives 
a smooth function
\begin{eqnarray}
\langle P({\bf Y}_{si},{\bf Y};t) \rangle = \frac{
\tau_{\rm D,S}^{-1} \, \cos \theta_{si} }
{4 (\Gamma_{\rm N} W_{\rm N}+\Gamma_{\rm S} W_{\rm S}) } \; 
\exp[-t/\tau_{\rm D,S}] \, ,
\label{eq:average-P}
\end{eqnarray}
where $W_{\rm N,S}=N_{\rm N,S} \pi/k_{\rm F}$ 
gives the width of the normal and superconducting leads.
We insert (\ref{eq:average-P}) into (\ref{eq:gamma-sum-to-Pintegral}),
and perform the integrals. Combining the result with Eqs.~(\ref{semicl-tr}),
(\ref{semicl-tr-spa}) and (\ref{eq:resummation})
delivers the resonant semiclassical contribution
to $T_{LL}^{he}$,
\begin{eqnarray}\label{eq:TLLhe}
\langle T_{LL}^{he} \rangle_{\rm r}  &=& \frac{\pi \Gamma_{\rm N}^2 N_{\rm N}}{4} 
\left(\frac{N_{\rm S}}{2 \Gamma_{\rm N} N_{\rm N} + 2 \Gamma_{\rm S} N_{\rm S}}
\right)^2 \\
& \times &\Big(1-(1+\tau_{\rm E}/\tau_{\rm D,S}) \exp[-\tau_{\rm E}/\tau_{\rm D,S}]\Big) \nonumber \\
& \times &\frac{\Gamma_{\rm S}}
{1-2 \, \Gamma_{\rm S} \, (1-\Gamma_{\rm N}) \cos[\pi-\phi]+\Gamma_{\rm S}^2 \, (1-\Gamma_{\rm N})^2} \, . \nonumber 
\end{eqnarray}

\begin{figure}[ht]
\begin{center}
\psfrag{g1} {\Large $\gamma_1$}
\psfrag{gn} {\Large $\gamma_n$}
\psfrag{g3} {\Large $\gamma_2$}
\psfrag{g2} {\Large $\gamma_3$}
\psfrag{R} {\Large $R$}
\psfrag{S1} {\Large $S_1$}
\psfrag{S2} {\Large $S_2$}
\psfrag{L} {\Large $L$}
\psfrag{ts1}{\Large $\theta_{s1}$}
\psfrag{ts2}{\Large $\theta_{s2}$}
\psfrag{t1}{\Large $\theta_{1}$}
\psfrag{-t1}{\Large $-\theta_{1}$}
\psfrag{t0}{\Large $\theta_{0}$}
\psfrag{-t0}{\Large $-\theta_{0}$}
\psfrag{y0}{\Large $y_{0}$}
\psfrag{ys1}{\Large $y_{s1}$}
\psfrag{ys2}{\Large $y_{s2}$}
\includegraphics[width=8.5cm]{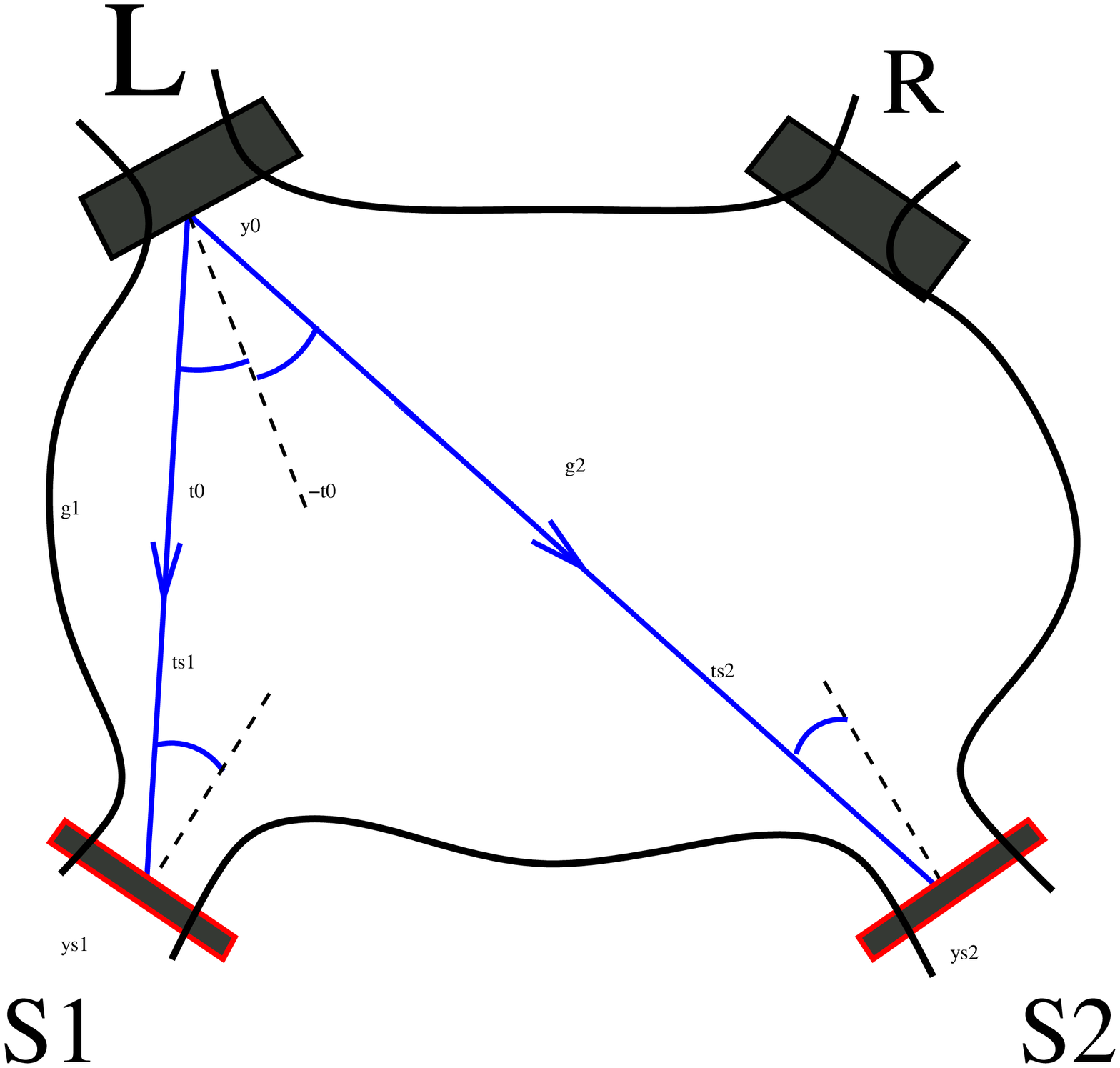}
\includegraphics[width=8.5cm]{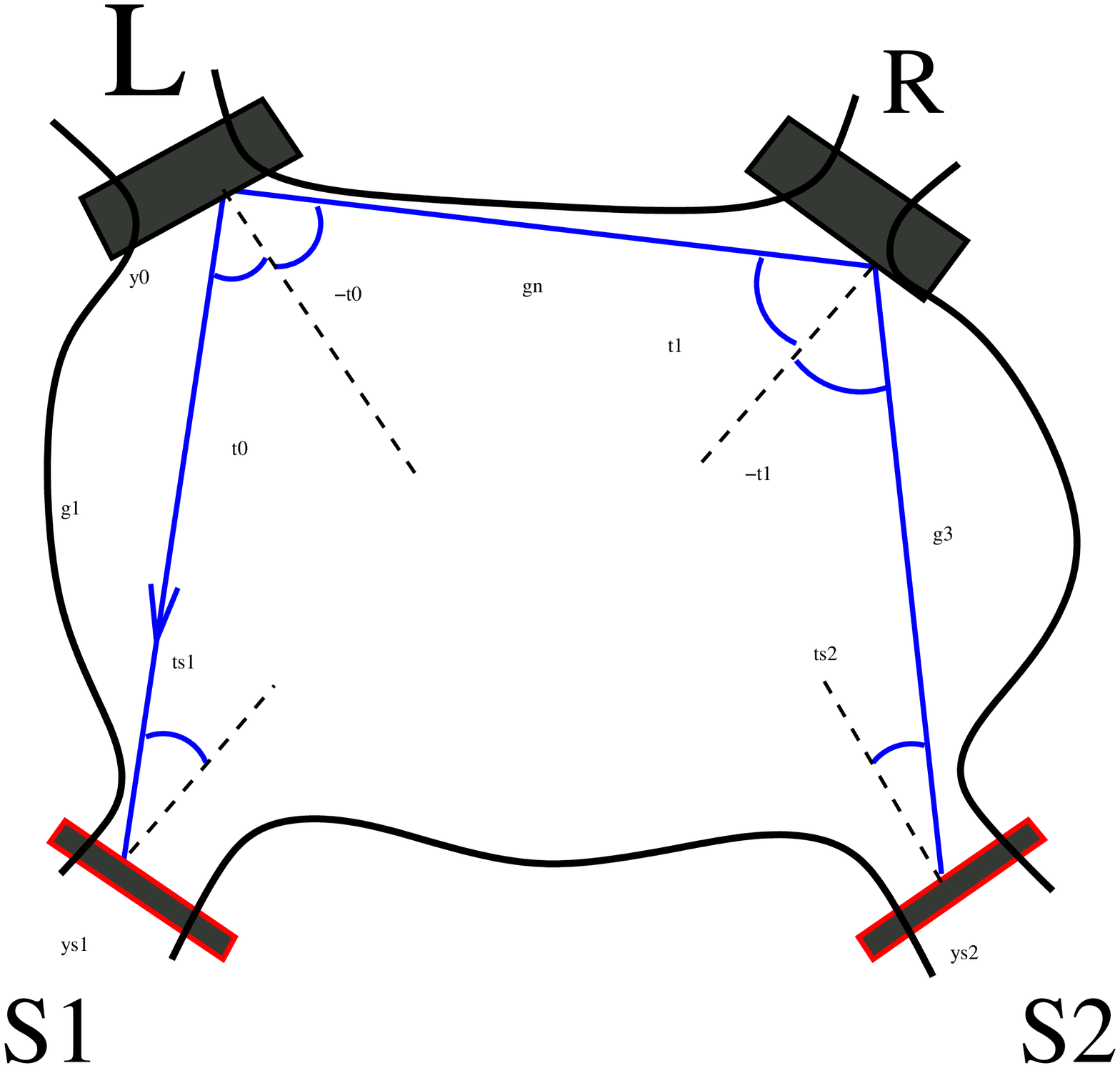}
\vspace{-1.1cm}
\end{center}
\caption{\label{fig:classI} The effective
four-terminal cavity used in the calculation of $\sum B_\gamma^2$,
and the geometry of class I (top panel) and class II (bottom panel) 
trajectories.}
\end{figure}

The semiclassical contribution to $T_{RL}^{ee}$ arises from
class II trajectories. The geometry is detailed in
Fig.~\ref{fig:classI}, and a similar calculation as above delivers
\begin{eqnarray} \label{eq:TRLee}
\langle T_{RL}^{ee}\rangle_{\rm r} 
&=& \frac{\pi^2 \Gamma_{\rm N}^2 \, N_{\rm N}^2}{8 
N_{\rm S}} \left(\frac{N_{\rm S}}{2 \Gamma_{\rm N} N_{\rm N} + 2 \Gamma_{\rm S} N_{\rm S}}\right)^3 \\
& \times &
\Big(1-(1+\tau_{\rm E}/\tau_{\rm D,S}+\tau_{\rm E}^2/2\tau_{\rm D,S}^2) 
\exp[-\tau_{\rm E}/\tau_{\rm D,S}]\Big) \nonumber \\
&\times& 
\frac{1+\Gamma_{\rm S}^2 \, (1-\Gamma_{\rm N})^2}{1-2 \, \Gamma_{\rm S} \, (1-\Gamma_{\rm N})^2 \cos[\pi-\phi]+\Gamma_{\rm S}^2 \, (1-\Gamma_{\rm N})^4}. \nonumber
\end{eqnarray}
The semiclassical contribution to the conductance 
is given by the sum of
Eqs.~(\ref{eq:TLLhe}) and (\ref{eq:TRLee}). It clearly exhibits the functional
dependence of resonant tunneling, where the resonance is however
always at the Fermi level, and is achieved by setting the phase
difference between the two superconductors at $\phi=\pi$. This resonance
condition is the same for all semiclassical contributions we 
just discussed. This is
why the resonance is macroscopic, giving a conductance 
$G \propto N_{\rm N}$ and not of order
one, as is usually the case  
for resonant tunneling. We also see that the effect
disappears if the superconductors are poorly connected to the normal cavity,
$\Gamma_{\rm S} \rightarrow 0$, as should be.

Comparing
Eqs.~(\ref{eq:TLLhe}) and (\ref{eq:TRLee}) to Eq.~(\ref{G12ex}) we see
that semiclassical contributions matter only close to resonance.
In the tunneling regime, 
they increase the conductance by a factor $\propto \Gamma_{\rm N}^{-1}$,
from $G(\pi)/G_0 \propto \Gamma_{\rm N} N_{\rm N}$ to 
$G(\pi)/G_0 \propto N_{\rm N}$, resulting in a peak-to-valley ratio of the
conductance $G(\phi=\pi)/G(\phi=0) \propto \Gamma_{\rm N}^{-2}$. In most 
instances, $\langle T_{LL}^{eh}(\pi)\rangle_{\rm r} \gg 
\langle T_{RL}^{ee}(\pi)\rangle_{\rm r} $, and the sharpness of the
resonance peak, measured by its width at half height,
is proportional to $\Gamma$. 

\section{Numerical simulations}
\label{sectionnum}
We compare our theory with numerical simulations on 
a quantum mechanical model of
the Andreev billiard, the Andreev kicked rotator~\cite{Jac03}. 

\subsection{The Andreev kicked rotator}

One-dimensional maps give a stroboscopic description of dynamical systems. An 
example is the kicked rotator, describing
the dynamics of a particle confined to a circle which is kicked
periodically in time with period $\tau_0$ 
and a kicking strength depending on its position
\cite{Sch05}. The quantized version of this map is characterized by its Floquet
operator, the unitary operator giving the time-evolution from one kick
to the next, i.e. from time $t$ to $t+\tau_0$. 
We represent the Floquet operator 
as a $M\times M$ matrix with matrix elements
(from now on we measure times in
units of $\tau_0 \equiv 1$) 
\begin{subequations}
\begin{eqnarray}
F_{nm}&=&e^{-(i\pi/2M)(n^2+m^2)}(UQU^\dag)_{nm},\\
U_{nm}&=&M^{-1/2}e^{(2\pi i/M)nm},\\
Q_{nm}&=&\delta_{nm}e^{-(iMK/2\pi)\cos(2\pi n/M)}.
\end{eqnarray}
\end{subequations}
Here $K$ is the kicking strength. For $K\gtrsim 7$ the classical dynamics
is chaotic, with classical Lyapunov exponent $\lambda \simeq \ln[K/2]$. 
The eigenvalues of $F$ are $e^{i\epsilon_m}$ ($m=1\hdots M$), and
the quasi-energies $\epsilon_m$ of the closed system
(without any leads) have an average level spacing $\delta=2\pi/M$.  
The dynamics of the electron is determined by 
$F$, while the time
evolution of holes is given by $F^*$.

The kicked rotator model describes a closed system. We open it
by inserting leads in a subspace of the Hilbert
space, parametrized by the indices $\{m_1,m_2,\ldots m_{2N_{\rm N}}\}$ for the normal 
leads and
$\{l_1,l_2, \ldots l_{2N_{\rm S}}\}$ for the superconducting leads. The coupling to 
the leads is
described by a $2N_{\rm N}\times M$ ($2N_{\rm S}\times M$) projection 
matrix $P_{\rm N}$
  ($P_{\rm S}$) with matrix elements
\begin{subequations}
\begin{align}
[P_{\rm N}]_{nm}&=\left\{\begin{array}{ll}
\sqrt{\Gamma_{\rm N}}& \mbox{if\,} m=n\in \{m_1, \hdots m_{2N_{\rm N}}\},\\
0 & \mbox {otherwise,}
\end{array}\right.\\
[P_{\rm S}]_{nm}&=\left\{\begin{array}{ll}
\sqrt{\Gamma_{\rm S}}& \mbox{if\,} m=n\in\{l_1, \hdots l_{2N_{\rm S}}\},\\
0 & \mbox {otherwise.}
\end{array}\right.
\end{align}
\end{subequations}
The average time between Andreev
 reflections equals $\tau_{\rm D,S}=M / 2\Gamma_{\rm S}N_{\rm S}$. In our numerical
investigations we restrict ourselves to $\Gamma_{\rm S} = 1$.
The Ehrenfest time of the closed Andreev billiard reads $\tau_{\rm E}=\lambda^{-1}\ln{(N_{\rm S}^2/M)}$~\cite{Jac03,Vav03}.

We perform two sets of simulations, the first is devoted to the spectrum
of a closed Andreev billiard, the second to the conductance through a
chaotic Andreev interferometer. 

\subsubsection{The closed Andreev kicked rotator}

Both electrons and holes have an
energy near the Fermi energy, 
and therefore we assume that Andreev reflection is perfectly retroreflecting.
The Floquet matrix of the Andreev billiard is then
\begin{align}
{\cal F}={\cal P}_S \, \left(
\begin{array}{cc}
F&0\\
0&F^*
\end{array}
\right) \, ,
\end{align}
with 
\begin{align}
{\cal P}_S=\left(
\begin{array}{cc}
1-P_S^{T}P_S & -iP_S^Te^{i\Phi}P_S\\
-iP_S^Te^{-i\Phi}P_S&1-P_S^TP_S
\end{array}
\right).
\end{align}
The coupling between electrons and holes 
is modelled by ${\cal P}_S$, 
which induces Andreev reflection for electron or holes touching the 
superconducting leads, i.e. induces transitions between the components 
$\{l_1, \ldots l_{2N_{\rm S}}\}$ of the wavefunction in the two
quasiparticle sectors.
The phase difference between the two superconductors is contained in
the $2N_{\rm S}\times 2N_{\rm S}$ matrix $\Phi$, with 
elements $\delta_{nm}\phi /2$, for $n\le N_{\rm S}$ and $-\delta_{nm}\phi /2$ 
for $n>N_{\rm S}$.  
The matrix ${\cal F}$ can be diagonalized efficiently using the Lanczos
technique in combination with the Fast-Fourier-Transform 
algorithm \cite{Ket99}.
This allows to reach the large system sizes needed to explore the
regime of finite Ehrenfest time.

\subsubsection{The kicked Andreev interferometer}
We first write the $(2N_{\rm N}+2N_{\rm S})\times (2N_{\rm N}+ 2N_{\rm S})$ energy-dependent scattering matrix of our four-terminal structure
(without superconductor). It is related to the Floquet operator 
via~\cite{Fyo00}
\begin{align}
S(\epsilon)=-(1-PP^T)^{1/2}+P\frac{1}{e^{-i\epsilon}-{F}(1-P^TP)}FP^T.
\end{align}
The $(2N_{\rm S}+2N_{\rm N})\times M$ coupling matrix 
\begin{align}
P=\left(\begin{array}{cc}
P_{\rm N}&0\\
0&P_{\rm S}
\end{array}\right),
\end{align}
describes coupling to all four leads.
The four-terminal 
scattering matrix $S$ is made of sixteen sub-blocks $S_{ij}$, where
the subindices $i,j=L$, $R$, $S1$ and $S2$ label either normal or
superconducting terminals. From these sub-blocks, the
exact expression for the conductance, Eq.~(\ref{eq:full-lambert}),
can be computed, following Ref.~\cite{Bee95}.
The transmission amplitudes giving the probabilities in 
Eq.~(\ref{eq:ampl_probs}) can be written as
\begin{subequations}\label{eq:trans_prob_eh}
\begin{eqnarray}
&&\tau^{\rm ee}=a-b\,\Omega c^{\ast}\Omega^{\ast}(1+c\,\Omega
c^{\ast}\Omega^{\ast})^{-1}d,\label{ree2}\\
&&\tau^{\rm he}=-{\rm i}b^{\ast}\Omega^{\ast}(1+c\,\Omega
c^{\ast}\Omega^{\ast})^{-1}d,\label{rhe2}
\end{eqnarray}
\end{subequations}
in terms of the matrices
\begin{eqnarray*}
&&a=
{\renewcommand{\arraystretch}{0.6}
\left(\begin{array}{cc}
S_{L,L}&S_{L,R}\\S_{R,L}&S_{R,R}
\end{array}\right)},\;\;\;\;\;\;\;\;\;
b=
{\renewcommand{\arraystretch}{0.6}
\left(\begin{array}{cc}
S_{L,S1}&S_{L,S2}\\S_{R,S1}&S_{R,S2}
\end{array}\right)},\nonumber\\
&&c=
{\renewcommand{\arraystretch}{0.6}
\left(\begin{array}{cc}
S_{S1,S1}&S_{S1,S2}\\S_{S2,S1}&S_{S2,S2}
\end{array}\right)},\;\;\;\;
d=
{\renewcommand{\arraystretch}{0.6}
\left(\begin{array}{cc}
S_{S1,L}&S_{S1,R}\\S_{S2,L}&S_{S2,R}
\end{array}\right)},\nonumber\\
&&\Omega=
{\renewcommand{\arraystretch}{0.6}
\left(\begin{array}{ll}
{\rm e}^{-{\rm i}\phi/2}&0\\0&{\rm e}^{{\rm i}\phi/2}
\end{array}\right)}.
\end{eqnarray*}
For numerical stability and efficiency, the transmission 
matrices $\tau$ of Eqs.~(\ref{eq:trans_prob_eh}) are 
constructed from self-consistent linear equations for the output 
amplitudes as a
function of unity input on each $N_{\rm N}$ channel, one at a time,
and not via direct inversion.

\begin{figure}
\includegraphics[width=4.6cm]{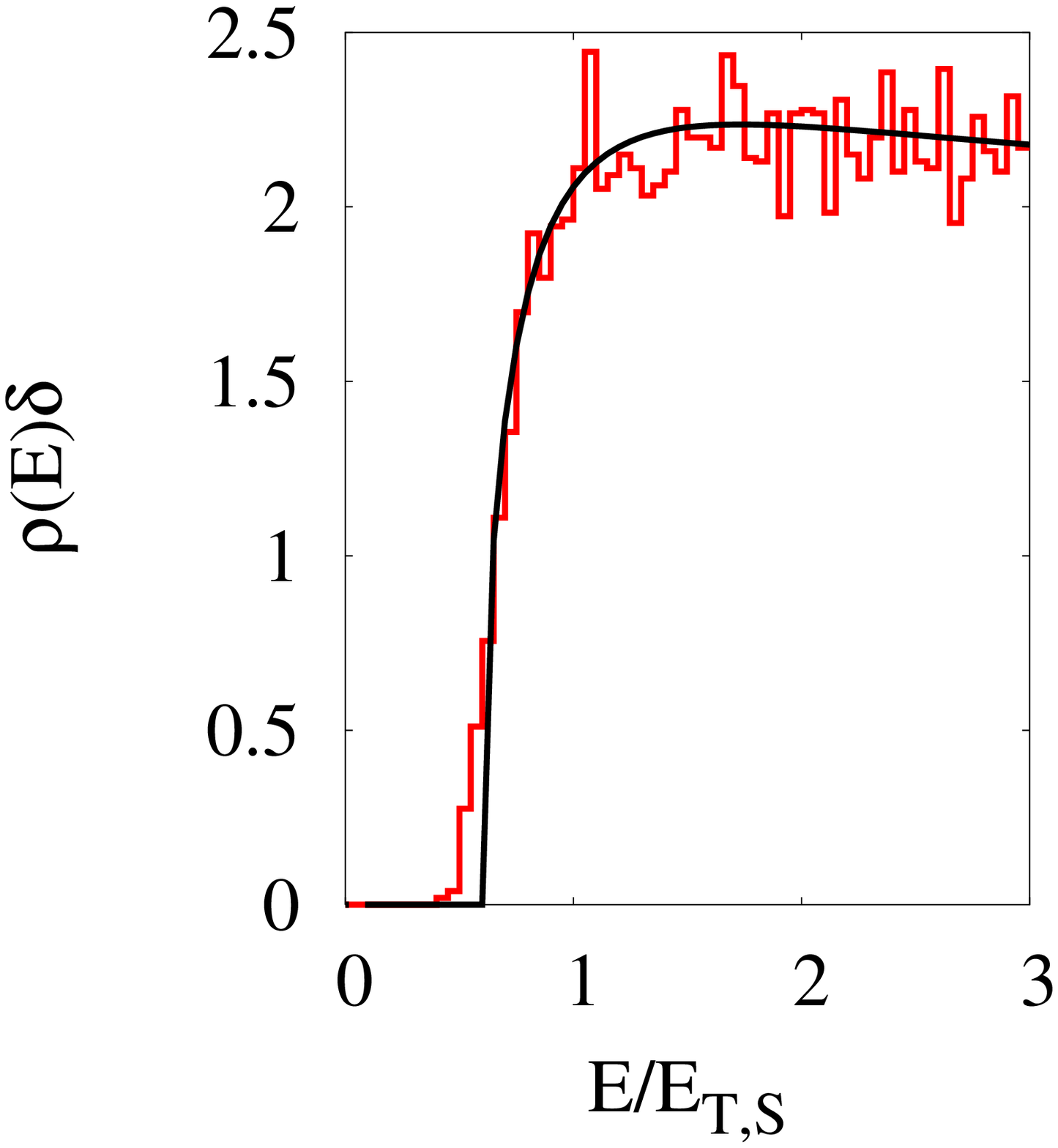}
\includegraphics[width=3.84cm]{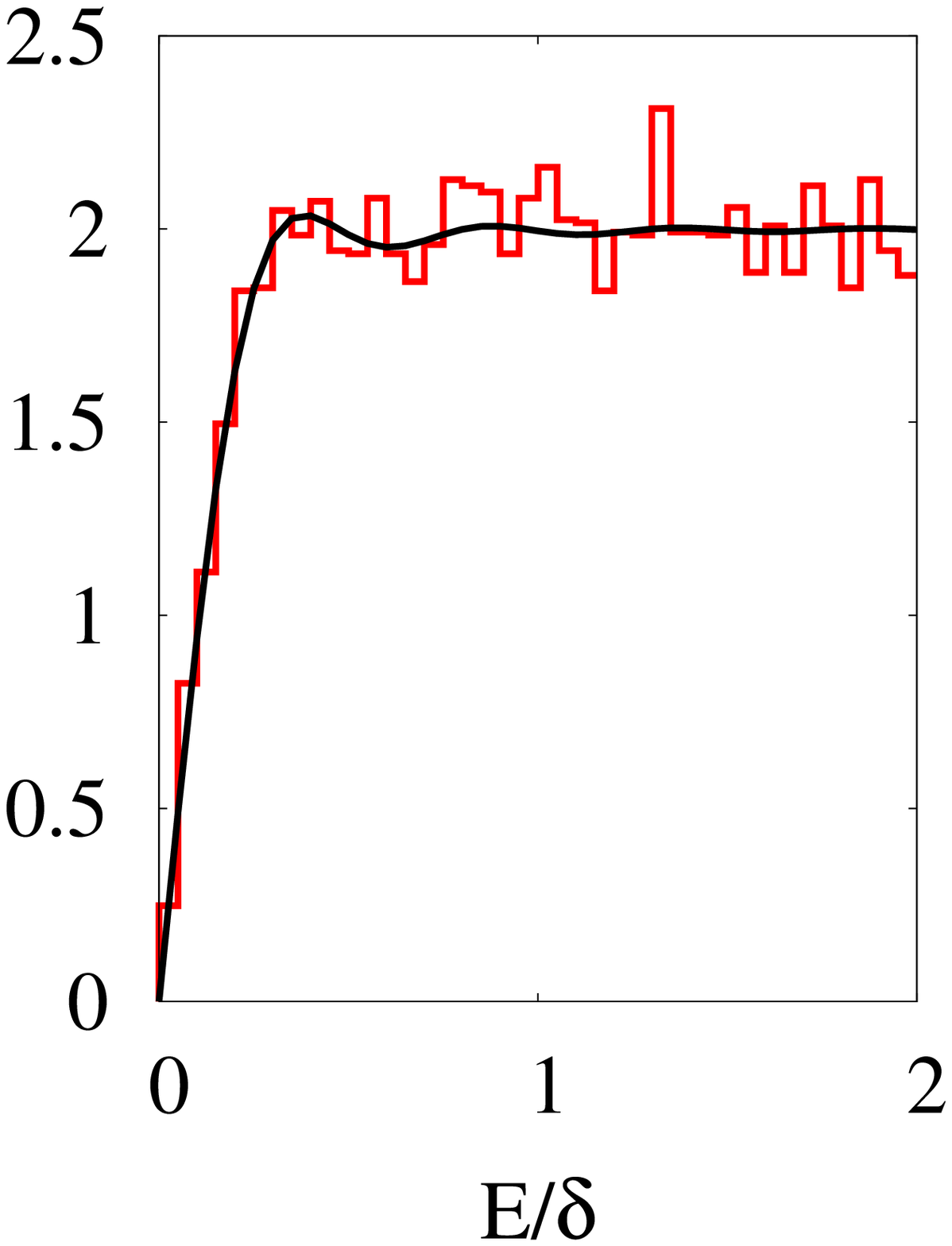}
\caption{\label{rhoRMTkick0}  Left panel:
density of states for the Andreev
kicked rotator with $\phi=0$,
$M=2048$, $N_{\rm S}=25$, $K=147$, $\Gamma_{\rm S}=1$.
The solid line gives the prediction of 
Eqs.\ (\ref{rhoRMT}) and (\ref{Greenfunction}). Data are 
averaged over $50$ different lead positions.
Right panel: density of states for the Andreev
kicked rotator with $\phi=\pi$,
$M=2048$, $N_{\rm S}=25$, $\Gamma_{\rm S}=1$.
The solid line gives the prediction of Eq.\ (\ref{rhozir}).
Data are averaged over $50$
different lead positions and $50$ values of the kicking strength 
$K \in [144.5 , 147.5]$}
\end{figure}

\subsection{Numerics in the universal regime}
We start with a comparison of the analytical predictions of RMT with
numerical results, by choosing the parameters of the  Andreev kicked rotator in such a way
that $\tau_{\rm E}\ll\tau_{\rm D,S}$.
In Fig.~\ref{rhoRMTkick0} we show the density of
states for $\phi=0$ and $\phi=\pi$. We compare
our results with the theoretical predictions of
Eqs. (\ref{rhoRMT}) and (\ref{Greenfunction}) for $\phi=0$, 
while for $\phi=\pi$ we use 
Eq.\ (\ref{rhozir}). In both cases we find perfect
agreement between numerics and RMT predictions. 

We next turn our attention to the conductance. We choose an Andreev billiard
with the same parameters as in Fig.~\ref{rhoRMTkick0}, but couple it to normal
leads. The conductance as a function of
the phase difference is shown
in Fig. \ref{GvsphiRMT}. The numerical results, for several
values of the parameters $\Gamma_{\rm N}$ and $N_{\rm N}$ (characterizing the normal leads),
are compared with the circuit theory prediction of Eqs.~(\ref{theta5}) and
~(\ref{G12}). The agreement is excellent. In the tunneling limit
$\Gamma_{\rm N}\rightarrow 0$ the conductance is strongly
suppressed for $\phi\ne \pi$ while it is restored to its value in
absence of superconductor when
$\phi=\pi$, as predicted by Eq.~(\ref{G12ex}).

\begin{figure}[ht]
\begin{center}
\includegraphics[width=8.5cm]{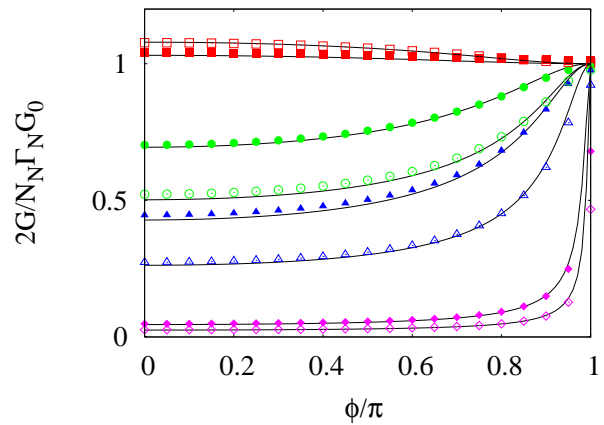}
\end{center}
\caption{\label{GvsphiRMT}  
Conductance of the Andreev kicked rotator as a 
function of the phase
  difference $\phi$ between the two superconductors. 
For all curves $M=2048$, $N_{\rm S}=25$, $K=147$. The closed symbols 
correspond to $N_{\rm N}=100$
and the open symbols to $N_{\rm N}=50$, with $\Gamma_{\rm N}=1$ 
(squares),
  $\Gamma_{\rm N}=0.2$ (circles), $\Gamma_{\rm N}=0.1$ 
(triangles) and $\Gamma_{\rm N}=0.01$
  (diamonds). The solid lines are the relevant predictions from circuit
  theory. Every data point is an average over $200$ realizations. }
\end{figure}

\begin{figure}[ht]
\begin{center}
\includegraphics[width=8.5cm]{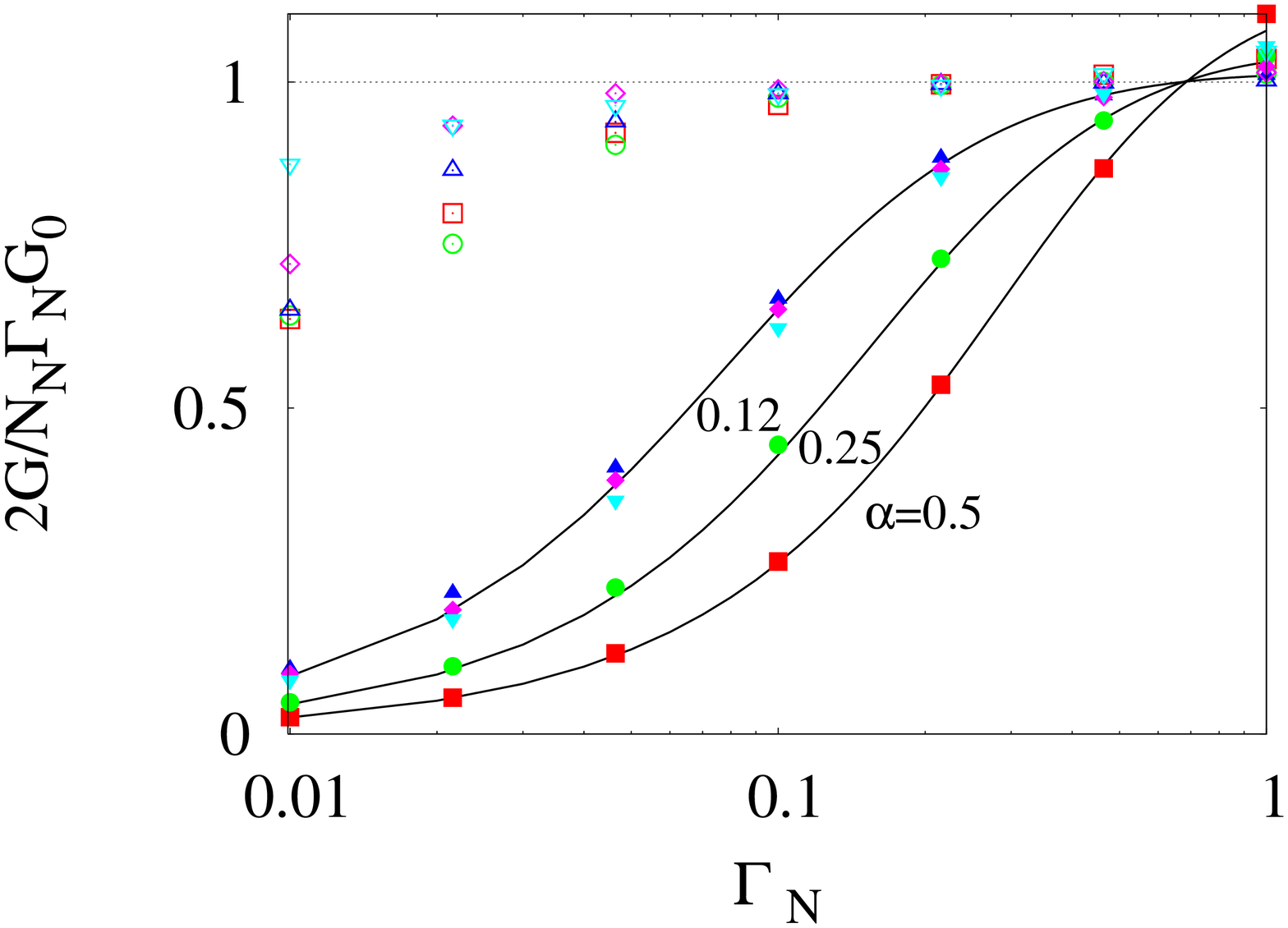}
\end{center}
\caption{\label{GvsGammaRMT}  Conductance of the Andreev kicked rotator as
  a function of $\Gamma_{\rm N}$ for $K=147$, $M=2048$, $\Gamma_{\rm S}=1$, 
at $\phi=0$ (closed symbols)
  and $\phi=\pi$ (open symbols), and  $N_{\rm N}/N_{\rm S}=100/50$
  (squares), $100/25$ (circles), $100/12$ (upward pointing triangles), $200/25$
  (diamonds) and $400/50$
  (downward pointing triangles). Every data point is an average over
  $50$ realizations.  
 The solid lines are the predictions from circuit theory
for $\phi=0$ and
  $\alpha=N_{\rm S}/N_{\rm N}=0.5, 0.25$, and $\alpha=0.12$, while
the dashed line corresponds to $\phi=\pi$. }
\end{figure}

In Fig.~\ref{GvsGammaRMT},
we show the conductance as a function of $\Gamma_{\rm N}$, 
both for $\phi=0$ and $\phi=\pi$. 
For $\phi=0$ the circuit theory
of Eqs.\ (\ref{theta5}) and (\ref{G12}) (solid lines) 
predict a curve that depends only on the
ratio $\alpha\equiv N_{\rm S}/N_{\rm N}$. The numerical data  
line up on the analytical curves. 
For $\phi=\pi$ the analytical prediction of circuit theory coincides with the
conductance of the billiard in the absence of superconductors i.e. $G/G_0=\Gamma_{\rm N}N_{\rm N}/2$. As 
$\Gamma_{\rm N} N_{\rm N}$ is reduced, however, the 
data deviate from this prediction. We attribute this to the fact that 
the circuit theory results are accurate
to leading order in $\Gamma_{\rm N} N_{\rm N}$ only.
This argument is corroborated by the fact that
the data for the largest $N_{\rm N}$ remain close to the
circuit theory prediction down to lower values of $\Gamma_{\rm N}$.

\begin{figure}
\begin{center}
\psfrag{j=p} {\color{red} \large $\phi = \pi$}
\psfrag{j=0} {\large $\phi = 0$}
\psfrag{Andreev gap} {Andreev gap}
\includegraphics[width=7.5cm]{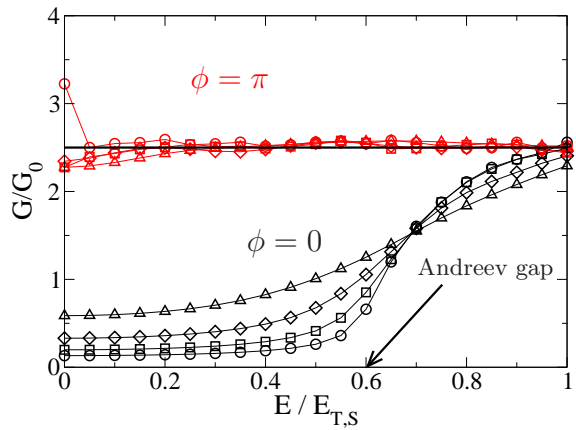}
\end{center}
\caption{\label{GvsGammaE}  Transmission as a function of
the normalized excitation energy $E/E_{\rm T,S}$ for $M=2048$, 
$N_{\rm N} = 100$, $K=147$, $\Gamma_{\rm N} = 0.05$, at $\phi=0$ (black
symbols) and $\phi=\pi$ (red symbols) for
$N_{\rm S}=25$ (triangles), 50 (diamonds), 100 (squares) and
200 (circles). The black curve shows the universal conductance
at $\phi=\pi$ for $E=0$.}
\end{figure}

We next present in Fig.~\ref{GvsGammaE} data for the conductance 
at finite excitation energy. For 
$\phi=\pi$, the conductance
stays at its universal, $E=0$ value,
$G = \Gamma_{\rm N} N_{\rm N}/2$, independent of $E$. For $\phi=0$
however, it increases from 
$G = \Gamma_{\rm N}^2 N_{\rm N} (1+2 N_{\rm N}/N_{\rm S})$
to $G = \Gamma_{\rm N} N_{\rm N}/2$ as $E$ increases. The crossover
occurs on the scale $\simeq 0.6 E_{\rm T,S}$ of the 
$\phi=0$ Andreev gap (see the top panel of Fig.~\ref{rhoRMTkick0}), which
illustrates further the connection between
transport through the Andreev interferometer and the spectrum
of the corresponding closed Andreev billiard. From the data in 
Fig.~\ref{GvsGammaE}, we do not
expect a fundamental departure from the zero-temperature theory 
in the universal regime, as long as the temperature lies below the
Thouless energy,
$T \lesssim E_{\rm T,S}$. The situation is different in 
the semiclassical regime, which can already be inferred from the
$N_{\rm S} = 200$ data presented in Fig.~\ref{GvsGammaE}. Despite
a rather small ratio $\tau_{\rm E}/\tau_{\rm D,S} \approx 0.13$, this set
of data already exhibits a peak at $E=0$
above the universal conductance
for $\phi = \pi$. This peak however quickly disappears 
with $E \ll E_{\rm T,S}$. 

We finally stress that for the data presented in Figs.~\ref{rhoRMTkick0}
\ref{GvsphiRMT} and
\ref{GvsGammaRMT},
$\tau_{\rm E}/ \tau_{\rm D,S}<0.02$. We found that,
to get a good agreement between our numerics and RMT
predictions at $\phi=\pi$, 
it is necessary to set $\tau_{\rm E}/ \tau_{\rm D,S}$ to much smaller values
than in our earlier numerical works with $\phi=0$~\cite{Jac03,Goo03,Goo05}.
This is especially true for the density of states at $\phi=\pi$ (where
we needed to go down to $\tau_{\rm E}/\tau_{\rm D,S}\approx 10^{-3}$), and
is already a good indication that the semiclassical
effect at $\phi=\pi$ is much larger than the manifestations of 
Ehrenfest physics investigated so far.

\subsection{Semiclassical limit}

We next investigate the 
spectrum of closed Andreev billiard at finite $\tau_{\rm E}/\tau_{\rm D,S}$.
Density of states of closed billiards 
are shown in Fig.~\ref{fig:DoS}. The density of states is only
weakly dependent on $\tau_{\rm E}/\tau_{\rm D,S}$ at $\phi=0$ 
(left panel).
At $\phi=\pi$ however, the semiclassical regime sees the emergence
of a giant peak in the density 
of states at the Fermi energy (right panel).

\begin{figure}
\begin{center}
\includegraphics[width=10cm]{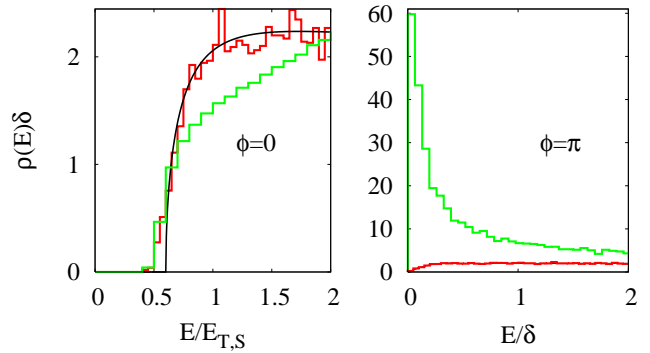}
\end{center}
\caption{
\label{fig:DoS}  Density of states of closed Andreev billiards,
for $\phi=0$ (left panel) and $\phi=\pi$ (right panel), in
the universal regime with
$M=2048$, $N_{\rm S}=25$ and $K=148$ (red curves, 
$\tau_{\rm E}/\tau_{\rm D,S}=0$, data averaged over
2500 samples) and in the semiclassical regime with 
$M=8192$, $N_{\rm S}=800$ and $K=10$ (green curves, 
$\tau_{\rm E}/\tau_{\rm D,S}=0.55$,
500 samples). The black curves indicate
the universal predictions~\cite{Mel96,Zir97}. Note the different horizontal
and vertical scales.\\[-0.65cm]
}
\end{figure}

\begin{figure}[ht]
\begin{center}
\psfrag{G(phi)} {$G(\phi)$}
\psfrag{G(pi)} {$G(\pi)$}
\psfrag{G(pi)/G(0)} {$\mbox{ \hspace{-0.5cm} $G(\pi)/G(0)$}$}
\psfrag{phi/2pi} {$\phi/2 \pi$}
\psfrag{kfl} {$M$}
\includegraphics[width=8.5cm]{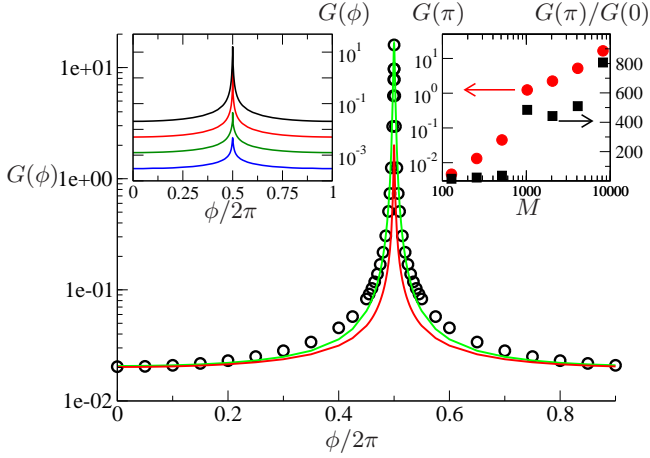}
\end{center}
\caption{
\label{fig:G_of_Phi}  
Conductance through a chaotic Andreev interferometer
vs. the phase difference $\phi$ between the two superconductors. 
Circles are numerical results obtained from the Andreev kicked 
rotator with $\Gamma_{\rm N}=0.01$, 
$M=8192$, $M/N_{\rm N}=20$, 
$M/N_{\rm S}=10$ and
$K=10$. The green curve is the analytical
prediction obtained by summing the semiclassical
resonant contributions with the
universal prediction. The red curve gives the
universal prediction obtained from circuit theory.
Left inset: Numerical data 
for the same classical parameters
$M/N_{\rm N}=20$, 
$M/N_{\rm S}=10$ and $K=10$ as in the main plot, for
$M=128$
(blue curve), 512 (green curve), 2048 (red curve), to
8192 (black curve). Note the change in peak-to-valley ratio.
Right inset: peak-to-valley ratio $G(\pi)/G(0)$ (black squares)
and peak conductance $G(\pi)$ (red circles) as a function of
$M$ controlling the quantum-to-classical
crossover, for the same classical configuration as in the main plot.
Data are averaged over 150--1000 sample realizations.\\[-8mm]}
\end{figure}

In Fig.~\ref{fig:G_of_Phi}, we next show a conductance 
resonance curve in the semiclassical
regime. We obtain very good agreement between
the numerical data (circles) and the analytical prediction (green solid line)
with $\tau_{\rm E}/\tau_{\rm D,S} \simeq 0.79$.
Without the semiclassical contribution, the universal 
prediction of Eqs.~(\ref{theta5}--\ref{G12}) in the full range 
$\phi \in [0,2\pi]$ is given by the red line. While the latter 
fits the resonance curve far away from resonance, the agreement breaks 
down close to $\phi=\pi$, where universal contributions give a prediction
$G(\pi)=2 G_0$, too small by an order of magnitude.
The left inset in Fig.~\ref{fig:G_of_Phi}
illustrates the increase of the peak height and narrowness as
the semiclassical parameter $M$ increases. The four sets of data in 
this inset correspond
to a fixed set of classical parameters, with the electronic wavelength 
decreasing by factors of four from one curve to the next, starting from
the bottommost (blue) curve. The conductance increases at each step because
the number of conduction channels scales linearly with $M$.
In absence of semiclassical contributions, these
four curves would exhibit the same peak-to-valley ratio, but here they
do not. This is 
quantified in the right inset to Fig.~\ref{fig:G_of_Phi},
where we show both the peak height and the peak-to-valley ratio
corresponding to the same classical configuration as in the main plot, 
while varying $M$.

We can connect the data in Figs.~\ref{fig:DoS} and \ref{fig:G_of_Phi},
since they correspond to the same closed Andreev billiard.
The broadening of levels due to the external normal leads is 
$\delta E \simeq \Gamma_{\rm N} N_{\rm N} \delta /2 \pi$.
For the resonance in Fig.~\ref{fig:G_of_Phi}, this gives
$\delta E =2 \delta/\pi$. From the green curve in Fig.~\ref{fig:DoS}, 
we estimate that this value of 
$\delta E$ covers about 20 to 30 levels around
the Fermi energy. Those states condense on classical orbits
bouncing on average $M/2 N_{\rm S}=5$ times
at the billiard's boundary. At each bounce, they have a probability
$\simeq 1-2 N_{\rm N}/(M-2 N_{\rm S})$ not to touch a normal lead.
Therefore, since
$[1-2 N_{\rm N}/(M-2 N_{\rm S})]^5 \approx 1/2$ for the data in
Fig.~\ref{fig:G_of_Phi} with $N_{\rm N}=400$ channels, we estimate that about
half of these 20--30 states couple to the leads. 
One thus expects a resonant
contribution to the conductance 
$\sim 10-15 G_0 $, which agrees quite well with the numerical value of
$G(\pi) \simeq 16 G_0$ for the data of Fig.~\ref{fig:G_of_Phi}, with a
universal contribution of $2 G_0$.
 
\begin{figure}[ht]
\begin{center}
\includegraphics[width=8.5cm]{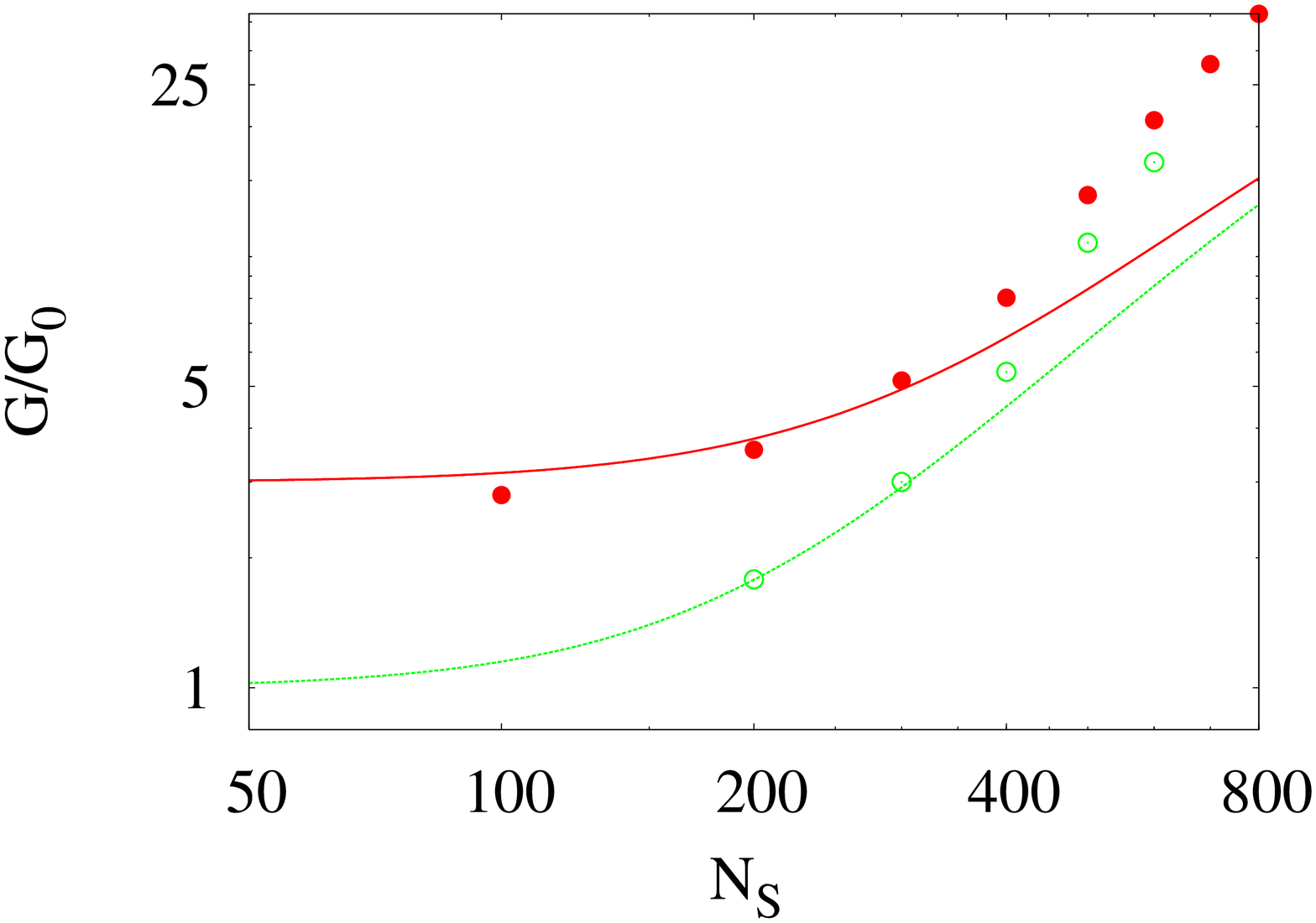}
\end{center}
\caption{\label{GvsNs} The conductance of the Andreev kicked rotator as
  a function of $N_{\rm S}$ for $K=10$, $M=4096$, $\Gamma_{\rm S}=1$, $\phi=\pi$, $N_{\rm
  N}=200$ and $\Gamma_{\rm N}=0.03$ (closed circles), $\Gamma_{\rm N}=0.01$ (open circles).
 Every data point is an average over at least
  $80$ realizations.  
 The
lines represent a sum of the semiclassical prediction Eq. (\ref{eq:TLLhe}) in the
  tunneling limit and the
  RMT result of Eq.\ (\ref{G12ex}) for $\Gamma_{\rm N}=0.03$
  (solid red line) and $\Gamma_{\rm N}=0.01$ (dashed green line). }
\end{figure}

In Fig.\ \ref{GvsNs} we show the conductance as a function of the number of
channels in each superconducting lead $N_{\rm S}$, for two different values of $\Gamma_{\rm N}$. We see that the
conductance depends strongly on $N_{\rm S}$, in contrast to the RMT prediction
$G=G_0\Gamma_{\rm N}N_{\rm N}/2$ which is independent of it. In the parameter range of the
figures, the contribution of Eq.\ (\ref{eq:TRLee}) can be neglected and the
contribution of Eq.\  (\ref{eq:TLLhe}) dominates. For $\phi=\pi$ and 
$\Gamma_{\rm N} \rightarrow 0$,
it can be simplified to $T_{LL}^{he}=(\pi N_{\rm N}/16)
\left(1-(1+\tau_{\rm E}/\tau_{\rm D,S}) \exp[-\tau_{\rm E}/\tau_{\rm
    D,S}]\right)$. This contribution depends strongly on $N_{\rm S}$ since 
both the Ehrenfest time and the time $\tau_{\rm D,S}$ 
between Andreev reflections depend on
$N_{\rm S}$. 
It is therefore the relative measure of trajectories of class ${\rm I}$
which causes the order-of-magnitude 
increase of conductance with $N_{\rm S}$ observed in Fig.~\ref{GvsNs}. 
The solid lines
give the theoretical prediction obtained by
summing the semiclassical resonant contribution, Eqs.~(\ref{eq:TLLhe})
and (\ref{eq:TRLee}), and the 
universal contribution obtained from Eqs.~(\ref{theta5}) and (\ref{G12}). 
There is good agreement between the numerical data and the theory, as
long as the superconduting dwell time is long enough. For 
$N_{\rm S} > 400$, $\tau_{\rm D,S} < 5$, and we suspect that short
processes across the billiard increase the conductance well above
the analytical prediction.

\begin{figure}[ht]
\begin{center}
\includegraphics[width=8.5cm]{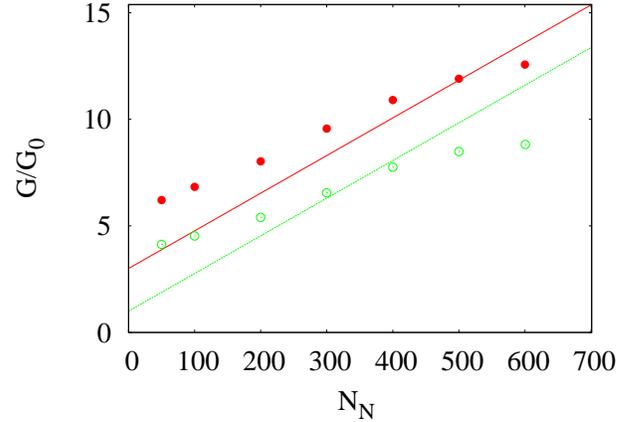}
\end{center}
\caption{\label{GvsNn} The conductance of the Andreev kicked rotator as
  a function of $N_{\rm N}$ for $K=10$, $M=4096$, $N_{\rm S}=400$, $\Gamma_{\rm S}=1$, $\phi=\pi$ and $\Gamma_{\rm N}=6/N_{\rm N}$ (closed circles), $\Gamma_{\rm N}=2/N_{\rm N}$ (open circles).
 Every data point is an average over at least
  $80$ realizations.  
 The
lines represent a sum of the semiclassical prediction Eq. (\ref{eq:TLLhe}) in the
  tunneling limit and the universal
  RMT result of Eq.\ (\ref{G12ex}) or $\Gamma_{\rm N}=6/N_{\rm N}$
  (solid red line) and $\Gamma_{\rm N}=2/N_{\rm N}$ (dashed green line). }
\end{figure}

Finally, we show on Fig.~\ref{GvsNn} 
the conductance as a function of $N_{\rm N}$. The product
$\Gamma_{\rm N}N_{\rm N}$ is kept constant, so that any 
observed $N_{\rm N}$ dependence arises from nonuniversal effects.
Even though we always have moderate Ehrenfest times
$\tau_{\rm E}<\tau_{\rm D,S}$, the numerical results are very different 
from the
constant conductance predicted in the universal regime.
The semiclassical contributions dominate
and we see an increase of $G$ with $N_{\rm N}$.
Here, $\tau_{\rm E}/\tau_{\rm D,S}\simeq 0.6$ is constant, and the increase
of the conductance is predicted to be linear, see
Eqs.\ (\ref{eq:TLLhe}) and (\ref{eq:TRLee}). There is reasonable 
agreement between theory and numerical data, with significant deviations
in the regime of small $N_{\rm N}$, where we leave the tunneling
regime.

We conclude that our numerical data fully confirm
our analytical results.

\section{Conclusion}
\label{sectioncon}

We have presented analytical and numerical investigations 
of spectroscopy and transport in quantum chaotic Andreev billiards coupled
to two superconductors.
We identified two regimes with very different behaviors. In the universal
regime, the presence of the superconductor opens up a gap in the
density of states of the Andreev billiard at the Fermi level, which
reduces the conductance through the system once it is attached to
two external normal leads with tunnel contacts. This gap closes when
the phase difference between the two superconductors is $\phi=\pi$,
and in the tunneling limit, 
the normal conductance that the system would have in the absence of
superconductivity is recovered. As the Fermi wavelength is reduced, 
$\lambda_{\rm F}/W_{\rm S}, \; \lambda_{\rm F}/L_{\rm c} \ll 1$, 
one enters a new, semiclassical
regime, where the situation is only marginally different unless the
phase difference between the two superconductors is close to $\pi$. 
When this is the case, we found an order-of-magnitude enhancement of the 
tunneling conductance, and identified
the mechanism behind this enhancement as resonant tunneling through
a macroscopic number of quasi-degenerate levels at the Fermi energy of
the corresponding closed Andreev billiard. 

Our calculation focused on the zero-temperature limit of
transport, and we finally comment on the effect of a finite
temperature. From Fig.~\ref{GvsGammaE},
we concluded that, in the universal regime, the shape of
the conductance oscillations do not differ from those presented in
Fig.~\ref{GvsphiRMT}, as long as the temperature lies below the superconducting
Thouless energy, $E_{\rm T,S}=\Gamma_{\rm S} N_{\rm S} \delta/2 \pi$.
The semiclassical contribution to the conductance 
near $\phi=\pi$ is however much more sensitive to changes in 
temperatures than the universal one. 
Finite excitation energies indeed lead to an additional
action phase accumulated along transport orbits. The expressions in
Eqs.~(\ref{eq:probabilities}) and (\ref{eq:resummation}) 
are now combined into
the following expression
\begin{eqnarray}
&&\frac{1}{\tau_{\rm D,S}^2}  \sum_{p,p' =0}^{\infty} (1-\Gamma_{\rm n})^{a(p+p')+b} \; \Gamma_{\rm s}^{p+p'+c} \; e^{i(p-p')(\phi-\pi)} \;\;\;\;\;\;\;\; \\
&& \times \int_0^{\tau_{\rm E}} {\rm d}t_{s1}
\int_0^{\tau_{\rm E}-t_{s1}} {\rm d}t_{s3} \;
e^{-[t_{s1}+t_{s3}][\tau_{\rm D,S}^{-1}+ i E (p-p')/\hbar]}.\nonumber 
\end{eqnarray}

The accumulation of an $E$-dependent action phase results
in a cut-off for the summation over $p$ and $p'$, effectively
suppressing contribution with $p-p' \gtrsim p_{\rm max}^{(-)}
\equiv (\hbar /  E \, \tau_{\rm D,S})^2$.
One ends up with a prefactor $1 - (1-\Gamma_{\rm N})^{p_{\rm max}^{(-)}}$
multiplying Eqs.~(\ref{eq:TLLhe}) and (\ref{eq:TRLee}), which
suppresses the semiclassical contribution already
for temperatures significantly smaller than
$E_{\rm T,S}$. At finite temperature, the resonant 
conductance enhancement discussed above will thus be
maximal for an intermediate value of $\Gamma_{\rm N}$ optimizing
macroscopic resonant tunneling while minimizing thermal averaging.\\

\section*{Acknowledgments}

We thank C. Beenakker for drawing our attention to Refs.~\cite{Kad95,Kad99},
M. B\"uttiker for helpful and valuable comments, and 
A. Kadigrobov for pointing out Ref.~\cite{Blom} and for a brief discussion on
Refs.~\cite{Blom,Kad95}.
M. Goorden was supported by the EU Marie Curie
RTN "Fundamentals of Nanoelectronics", MCRTN-CT-2003-504574.
P. Jacquod expresses his gratitude to M. B\"uttiker and the
Department of Theoretical Physics at the University of Geneva for their
hospitality during the summer of 2007.


\begin{thebibliography}{99}
\bibitem{caveat} The segments $\gamma_{si}$, $i=1,2,3$ in 
Fig.\ref{fig:device} might include many intermediate
bounces at a tunnel barrier. All possible number of reflections are 
effectively resummed; see Ref.~\cite{Whi07} for more details.
\end{thebibliography}
\end{document}